\documentclass[aps,prb,showpacs,reprint,superscriptaddress,twocolumn]{revtex4-2}

\usepackage{graphicx}
\usepackage{color}
\usepackage{amsmath}
\usepackage{bbm}
\usepackage{amssymb}

\usepackage{dsfont}

\usepackage[utf8]{inputenc}	
\usepackage[T1]{fontenc}

\usepackage[svgnames]{xcolor}

\input epsf

\begin{document}

\title{Charge and spin properties of a generalized Wigner crystal realized in the moiré WSe$_2$/WS$_2$ heterobilayer}
\author{Andrzej Biborski}
\email{andrzej.biborski@agh.edu.pl}
\affiliation{Academic Centre for Materials and Nanotechnology, AGH University of Krakow, Al. Mickiewicza 30, 30-059 Krakow,
Poland}
\author{Micha{\l} Zegrodnik}
\email{michal.zegrodnik@agh.edu.pl}
\affiliation{Academic Centre for Materials and Nanotechnology, AGH University of Krakow, Al. Mickiewicza 30, 30-059 Krakow,
Poland}

\begin{abstract}
We examine the charge and spin properties of an effective single-band model representing a moiré superlattice of the WSe$_{2}$/WS$_{2}$ heterobilayer. We focus on the $2/3$ electron filling, which refers to the formation of a generalized Wigner crystal, as evidenced experimentally. Our approach is based on the extended-Hubbard model on a triangular lattice with non-interacting part effectively describing a spin-split band due to Ising-type spin-orbit coupling. We investigate the system in the regime of strong on-site Coulomb repulsion and the ground state of the Hamiltonian is obtained with the use of the Density Matrix Renormalization Group formulated within the Matrix Product State approach. According to our analysis, on the basis of the density-density correlation functions resolved in the momentum space, a transition from the metallic to the insulating state appears with increasing intersite electron-electron interactions. This transition is identified as being concomitant with the emergence of a generalized Wigner crystal that realizes the honeycomb lattice pattern. We investigate the magnetic properties of such a Wigner crystal state and find that the increased intersite repulsion results in out-of-plane antiferromagnetic correlations and in-plane canted spin-spin ferromagnetic correlations. The latter are shown to become antiferromagnetic when the ground state is subjected to the transform, which simultaneously converts Hamiltonian into the SU(2) invariant form. 
\end{abstract}

\maketitle

\section{Introduction}
Strong electronic correlations in fermionic systems are believed to be the origin of a number of exotic phenomena that cannot be understood with the help of single-particle or mean-field theories. The widely known examples of such effects are: the formation of the insulating gap induced by local Coulomb repulsion which is the essence of the Mott-Hubbard transition~\cite{Imada1998,Cao2018_1}, variety of spin- or charge-ordered phases~\cite{Imada1998}, as well as the pairing mechanism resulting in unconventional superconductivity~~\cite{Keimer2015,Cao2018_2}. 

The role of electron-electron interactions in many particle systems is not limited to typical condensed matter systems, i.e. to those where electronic degrees of freedom are coupled to the underlying ionic lattice structure. In fact, the formation of a Wigner crystal~\cite{Wigner1934} (WC) in the electron gas can be considered a canonical example of correlation-induced effects when interactions between electrons predominate the kinetic energy contribution at the properly selected particle concentration~\cite{Chiao2024}. The idea of WC formation is almost one hundred years old~\cite{Chiao2024}. Over the years the pursuit for its experimental realization has motivated the study of 2D systems in the presence of relatively high magnetic fields, which suppress the contribution resulting from the kinetic energy due to the formation of Landau levels. Very recently, a direct visualization of WC by high-resolution scanning microscopy has allowed identification of its symmetry and melting in Bernal-stacked bilayer graphene~~\cite{Tsui2024}. The recent discovery of flat bands in moiré transition metal dichalcogenide bilayers, in particular WSe$_2$/WS$_2$ heterobilayer, has opened a new route for the realization of WC without the need to apply a magnetic field~\cite{Regan2020,Xu2020,Huang2021,Li2021,Zhou2024}. In fact, in such a system, the obtained pattern of localized charges should be considered as a generalized Wigner (GWC) crystal, since it is realized in an environment created by the underlying crystal lattice as pointed out by Hubbard\cite{Hubbard1978}. What distinguishes GWC from the \emph{canonical}       \cite{Wigner1934} WC is that in GWC spatial degrees of freedom of the particles to move are reduced by the periodic lattice potential. At selected fractional fillings of the WSe$_2$/WS$_2$ flat band, a GWC state can be created that may be considered as a \emph{extreme} form of charge density wave (CDW) induced by the non-local (intersite) Coulomb interactions. In such a scenario, only certain sublattices of the moiré superlattice are occupied, and a gap opens as suggested by the experiments~\cite{Xu2020,Huang2021}. In this view the GWC is encoded not only in the sharp redistribution of particle density with respect to sublattices but also in the opening of the gap, which can be considered as the so-called \emph{Wigner-Mott} transition\cite{Musser2022,Musser2}. Thus, the GWC can be identified when all the above-mentioned circumstances arise.

The moiré flat band of WSe$_2$/WS$_2$ is believed to be able to be described using an effective single band model~\cite{Wu2018} on a triangular lattice with spin-valley locking incorporated with the use of the spin-dependent complex phase of the hoppings~~\cite{Rademaker2022}. Supplementing such an approach with the electron-electron interaction terms, one obtains the Hubbard-like description~~\cite{Tan2023}. Along these lines the formation of the GWC realizing a Kagome pattern at $n=3/4$ filling, as well as its magnetic properties have been analyzed in the strong coupling limit when the hopping processes are completely suppressed, leading to a Heisenberg model supplemented with the Dzyaloshinskii-Moriya term~~\cite{Motruk2023}. Also, interesting studies of the physics of GWC formation on a triangular lattice with the use of extended Hubbard and continuum models have been reported in Refs. \onlinecite{Watanabe2005,Tan2023,Potasz,Ung2023,Amaricci2010,Musser2022, Musser2}, however without the inclusion of spin-valley locking. 

Here, we focus on analyzing the charge pattern formation leading to the GWC as well as its magnetic properties at selected fractional filling ($n=\frac{2}{3}$) by taking into account both the strong electron correlation effects induced by long-range Coulomb interactions and the spin-valley locking, at the same time. We start from establishing the model and briefly present the applied computational method, which is the Density Matrix Renormalization Group (DMRG) in the Matrix Product State (MPS) formulation. The DMRG method is considered to be one of the \emph{state of art} approaches for interacting fermionic systems. Next, we present both charge and spin properties by investigating one- and two-particle correlation functions. Treating the nearest-neighbor (nn) intersite Coloumb interaction $V$ as a free parameter, we demonstrate the indicators that reveal the transition from the metallic state to the insulating state for which the GWC emerges concomitantly with increasing $V$. Also, by inspecting the spin-spin correlation function we provide evidence of coexisting canted antiferromagnetism for the values of $V$, which recovers experimental findings. To the best of our knowledge, this kind of analysis is performed for the first time. We conclude our findings in the last Section of the paper.


\section{Model and method}
\begin{figure}
    \centering
    \includegraphics[width=0.5\linewidth]{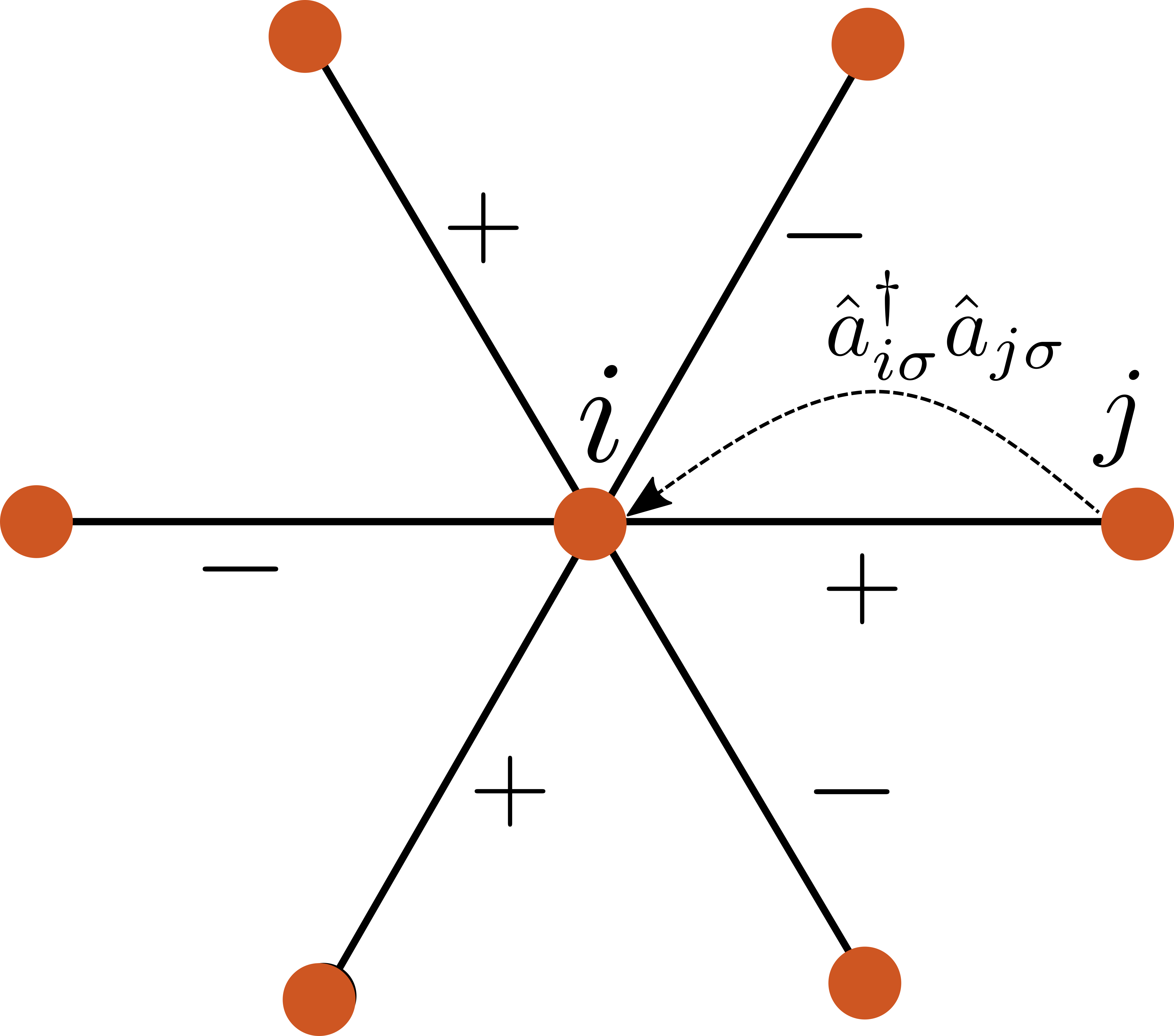}
    \caption{Schematic representation of  the sign of phase  $\phi_{ij}$ from Eq. \ref{eq:H0}. The dashed arrow represents exemplary hopping process which is weighted by factor $|t|\text{e}^{i\Tilde{\sigma}\phi_{ij}}$.}
    \label{fig:hopping_scheme}
\end{figure}

\begin{figure}
    \centering
    \includegraphics[width=0.95\linewidth]{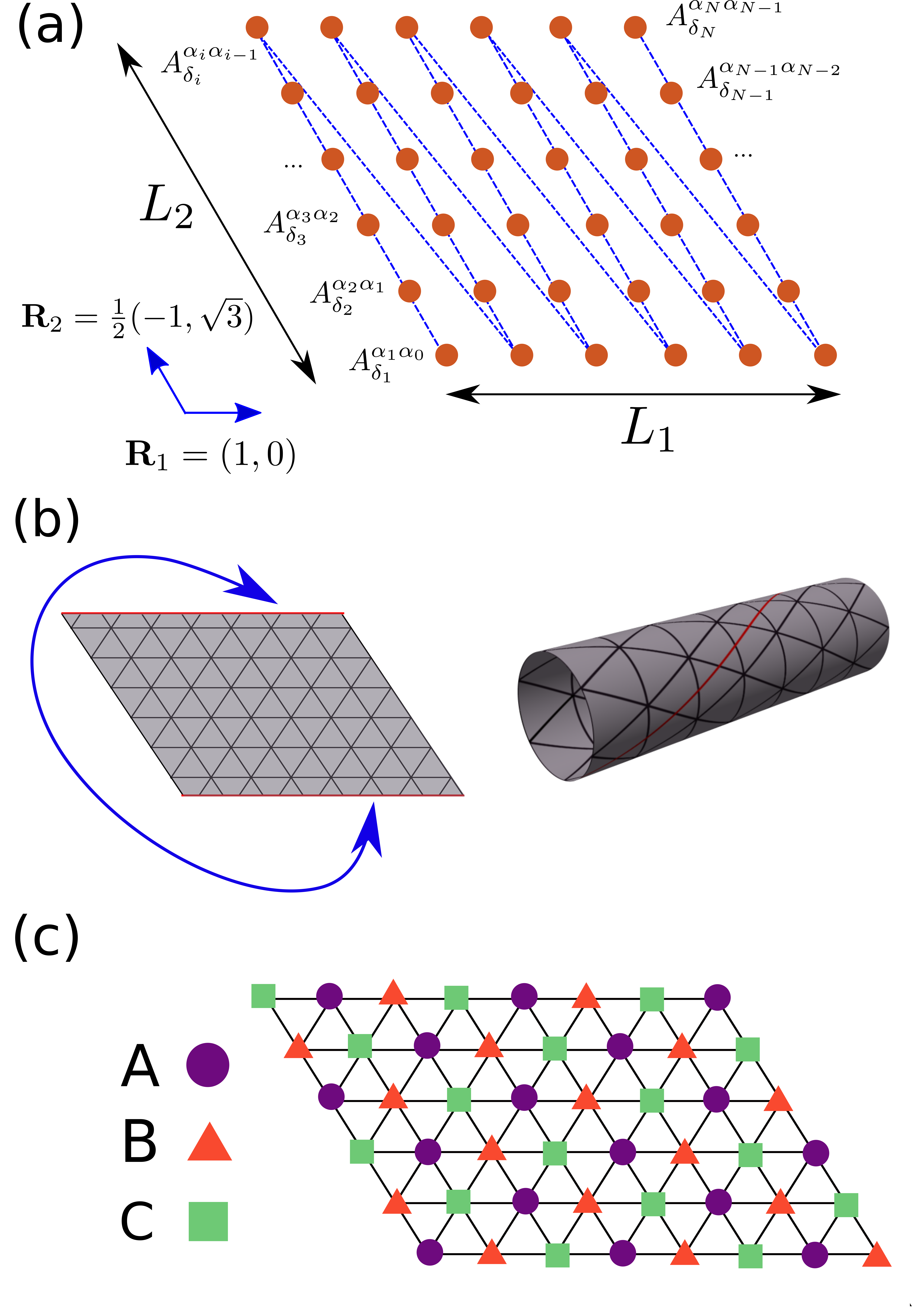}
    \caption{(a) The sketch of  MPS where the chosen ordering of $A$ matrices related to a given lattice sites is optimal for providing periodic boundary conditions along $\mathbf{R}_2$ vector. The total number of lattice sites is $N=L_1\times L_2$ where $L_1$ and $L_2$ are numbers of unit cells in directions given by $\mathbf{R}_1$ and $\mathbf{R}_2$, respectively. (b) The periodic boundary conditions are provided along $\mathbf{R}_2$. The red lines in 2D lattice supercell are \emph{glued} together creating a cylinder shape provided in the Figure. (c) The triangular lattice can be divided into three triangular sublattices which are labeled by $A$, $B$ and $C$. }
    \label{fig:mps}
\end{figure}

The single particle part of the Hamiltonian effectively describing spin-split flat moiré band of WSe$_2$/WS$_2$ is given as~\cite{Rademaker2022}
\begin{align}
    \mathcal{\hat{H}}_0=\sum_{\langle i,j\rangle}\sum_{\sigma}|t|\text{e}^{i\Tilde{\sigma}\phi_{ij}}\hat{a}^{\dagger}_{i,\sigma}\hat{a}_{j,\sigma},
    \label{eq:H0}
\end{align}
where $\hat{a}^{\dagger}_{i\sigma}(\hat{a}_{i\sigma})$ are standard fermionic operators creating (anihilating) electrons with spin z-component $\sigma=\{\uparrow,\downarrow\}$ at lattice site $i$. The spin dependent complex hoppings in $\mathcal{\hat{H}}_0$ incorporate the Ising type spin-orbit coupling that appears in the considered system and results in a spin-valley locking. In our notation $\tilde\sigma=1(-1)$ for $\sigma=\uparrow(\downarrow)$ and $\phi_{ij}=\pm\frac{2}{3}\pi$  with the sign convention depicted in Fig. \ref{fig:hopping_scheme}.
The interacting part of the Hamiltonian is taken as
\begin{align}
    \mathcal{\hat{H}}_{e-e}=\sum_{i}U\hat{n}_{i\uparrow}\hat{n}_{i\downarrow}+\frac{1}{2}\sum_{\langle ij\rangle}V\hat{n}_{i}\hat{n}_{j},
    \label{eq:He-e}
\end{align}
where $\hat{n}_{i\sigma}\equiv\hat{a}^{\dagger}_{i\sigma}\hat{a}_{i\sigma}$ and $\hat{n}_i\equiv\hat{n}_{i\uparrow}+\hat{n}_{i\downarrow}$. The first and the second term in Eq. (\ref{eq:He-e}) describes the onsite and nearest neighbor (nn) intersite Coulomb repulsion, respectively. The resulting model consists of both (\ref{eq:H0}) and (\ref{eq:He-e}) terms ,
\begin{align}
    \mathcal{\hat{H}}=\mathcal{\hat{H}}_0+\mathcal{{\hat{H}}}_{e-e},
    \label{eq:Htot}
\end{align}
and represents an extended Hubbard model (EH) on a triangular lattice with spin-valley locking.
To characterize the approximate many-particle ground state $|\Psi\rangle$ of Hamiltonian (\ref{eq:Htot}) we employ finite DMRG~\cite{White1992} approach in MPS formulation~\cite{SCHOLLWOCK201196,Catarina2023} using  ITensor library~\cite{itensor1}. In this variant, both the \emph{ansatze} of $|\Psi\rangle$ and the Hamiltonian at hand are given by products of matrices in such manner that the DMRG sweeping procedure can be efficiently implemented taking advantage of the tensorial nature of these representations of states and operators~\cite{SCHOLLWOCK201196}. The resulting MPS for the system consisting of the $N$ lattice sites is given as~\cite{SCHOLLWOCK201196}
\begin{align}
    |\Psi\rangle= \sum\limits_{\substack{ \delta_1,\cdots\delta_N \\ \alpha_1,\cdots,\alpha_{N-1}}}
    A_{\delta_1}^{\alpha_1\alpha_0}A_{\delta_2}^{\alpha_2\alpha_1}\cdots A_{\delta_N}^{\alpha_N\alpha_{N-1}}|\delta_1,\delta_2,\cdots\delta_N\rangle,
    \label{eq:MPS}
\end{align}
where $\{\delta_{i}\}$ refer  to physical degrees of freedom at $i$-th lattice site  and entries of matrices $A_{\delta_{i}}^{\alpha_i\alpha_{i-1}}$ are to be determined.
Although MPS can be considered as the natural way to describe the ground state of one-dimensional Hamiltonians, it can also be adapted to two-dimensional (2D) cases. To achieve this, we apply the ordering of the lattice sites (see Fig. \ref{fig:mps}a) to emulate the 2D system with the vectors $\mathbf{R}_1=(1,0)$ and $\mathbf{R}_2=\frac{1}{2}(-1,\sqrt{3})$.  In addition, we impose the periodic boundary conditions along the $\mathbf{R}_2$ direction and open boundary conditions in the $\mathbf{R}_1$ direction as presented in Fig. \ref{fig:mps}(b). The Hamiltonian represented by the matrix product operator (MPO) has complex-valued entries (as well as the MPS). Therefore, the operations during DMRG sweeps are more demanding than in the real-valued case. Eventually, the supercell considered with the boundary conditions imposed refers to the cylinder, which is a standard approach to investigate 2D systems in the framework of DMRG methods~\cite{Shirakawa2017,Szasz2020,Szasz2021}.


\section{Results}

All calculations discussed here correspond to $|t|=1$ and $U=15|t|$ representing a strongly correlated scenario that is believed to be realized in WSe$_2$/WS$_2$ heterobilayer. However, this choice of $U$ can be considered as close to the lower-bound approximation. That is, there is no consensus on the amplitude of interactions in this system and estimates differ by order of magnitude, since $U/|t|\approx
10\sim100$ \cite{Tang2020,Tan2023,Tang2023}. 
The intersite repulsions are, on the other hand, estimated as $U/V\approx4\sim5$~[\onlinecite{Wu2018,Tan2023}]. We extend our analysis to $V\in(0,6|t|)$ in order to study the gradual formation of the GWC through a precusor CDW state as the strength of the intersite Coulomb term is increased. We set $L_1=24$ and $L_2=6$ in the cylindrical system, since the order of the emerging lattice structure of the GWC (for significant values of $V$) is expected to be commensurate~\cite{Tocchio2014,Tan2023} with the period $\Delta L_{1,2}=3$ for band filling $n=N_e/N=\frac{2}{3}$, where $N=L_1\times L_2$, and,  $N_e$ is a number of carriers considered. In addition to imposing a constant number of particles, we also assume a total z-spin component $S_{tot}^{z}=\sum_i\langle \hat{S}_i^z\rangle=0$ (see the appendix for more details). All calculations have been performed for the maximal link dimensions [$\alpha_i$ in Eq. (\ref{eq:MPS})] up to 8192, resulting in a truncation error lower than $10^{-4}$ for all the cases considered. Also, based on the estimation of errors of the computed quantities (see the Appendix for details), we find the resultant data as reliable in view of the performed interpretation.

\subsection{Charge order and gap}
\begin{figure}
    \centering
    \includegraphics[width=1.0\linewidth]{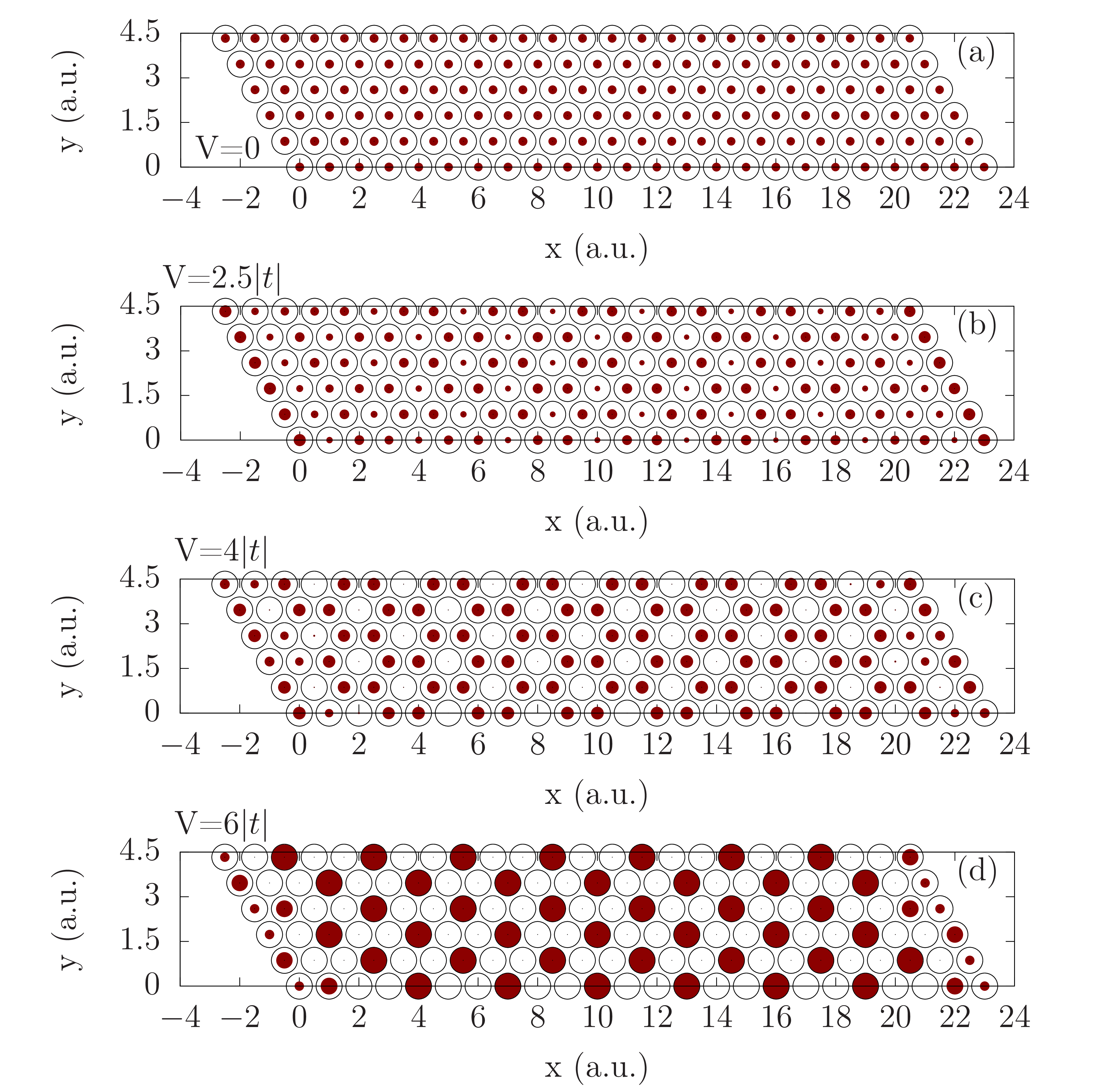}
    \caption{Spatial charge distribution in $24\times6$ cylinder for the three representative values of $V$. The hollow circles represent the lattice sites. The diameter of the filled red circles is proportional to $\langle\hat{n}_{i(x,y)}\rangle$. The completely filled circles represent doubly occupied sites.}
    \label{fig:nocc}
\end{figure}

We first determine the influence of the term $V$ on the spatial distribution of the charge over the moiré lattice sites. In Fig. \ref{fig:nocc}, we present $\langle\hat{n}_{i(x,y)}\rangle$ for the representative values of $V$. The complete set of data collected for various values $V$ is available in an online repository~\cite{andrzej_biborski_zenodo}. As can be seen, the electron distribution is almost homogeneous (disregarding minor edge effects) for $V=0$ [Fig. \ref{fig:nocc}(a)], whereas with increasing $V$ the CDW develops [Fig. \ref{fig:nocc}(b)]. Subsequently, for $V\gtrsim3|t|$ a clear honeycomb pattern of occupied lattice sites can be observed with all remaining sites, forming a triangular sublattice, being empty, as shown in Fig. \ref{fig:nocc}(c). For even stronger intersite Coulomb strength, when $V\approx6.0|t|$, a reversed situation occurs with a triangular sublattice occupied (Fig. \ref{fig:nocc}d) and a honeycomb sublattice empty. 
This behavior is in agreement with the one corresponding to the case of EH in a triangular lattice obtained without the inclusion of spin-valley locking ~\cite{Watanabe2005,Tocchio2014}. Following Tocchio et al.~\cite{Tocchio2014} we employ the nomenclature with respect to the three recognized types of order, e.g. $(\frac{2}{3}\frac{2}{3}\frac{2}{3})$, $(110)$ and $(200)$ referring to the situation where carriers occupy three, two and one triangular sublattice (-s), each labeled by $\alpha\in\{A,B,C\}$ (see Fig.\ref{fig:mps}c).  

\begin{figure}
    \centering
    \includegraphics[width=1.0\linewidth]{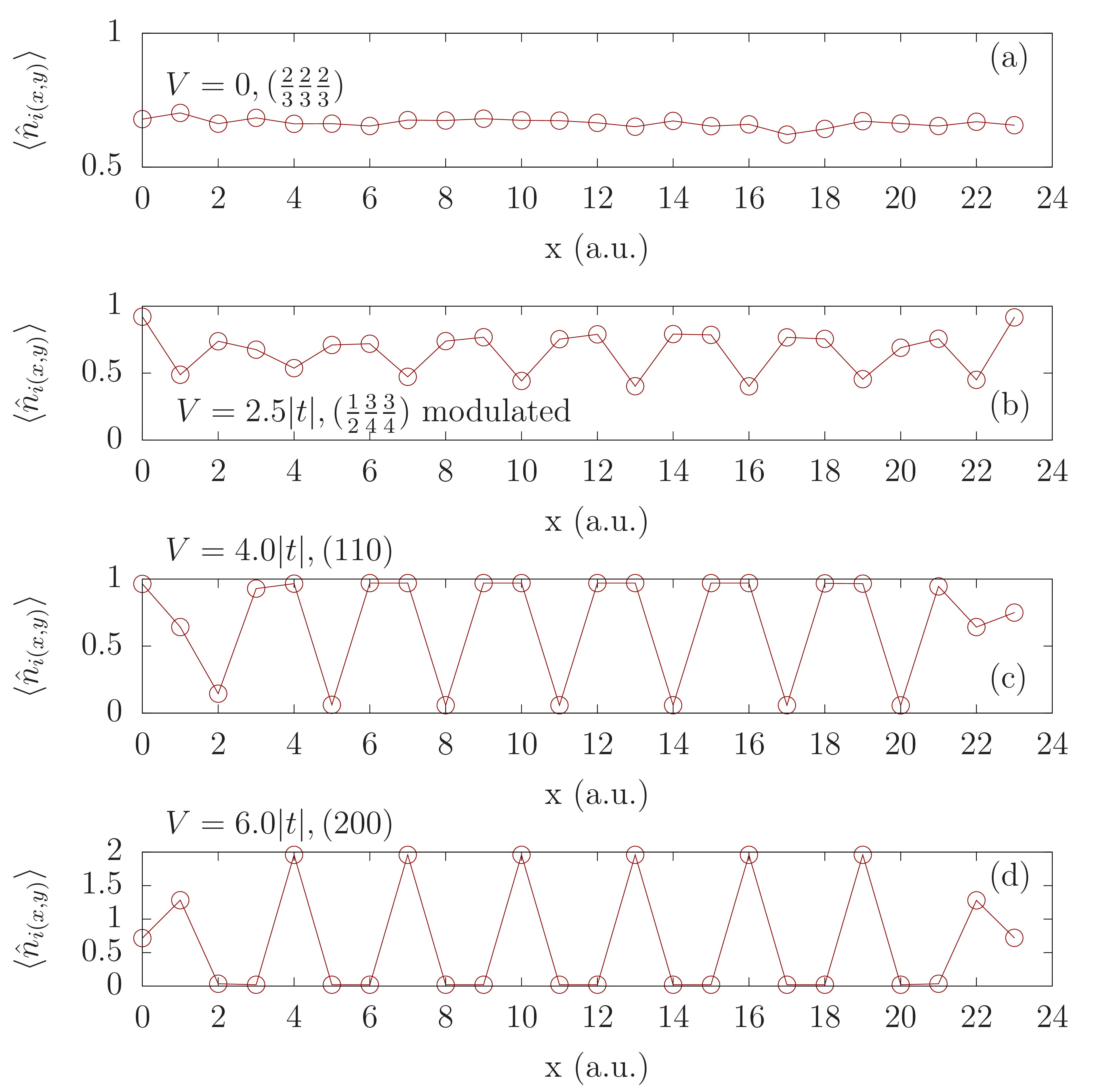}
    \caption{Value of $\langle\hat{n}_{i(x,y=0)}\rangle$ for the $24\times 6$ system for the the representative values of $V$. Excluding the situation $V=0$ (a), the modulation by a period of $\Delta L=3$ is clearly visible.}
    \label{fig:noccy0}
\end{figure}

In Fig. \ref{fig:noccy0} we present our results in a more quantitative way, that is, we plot the electron occupation at subsequent lattice sites along the direction $\mathbf{R}_1$ at $y=0$. As can be seen, the finite size of the system and the open boundary conditions in the $\mathbf{R}_1$ direction only slightly affect the densities close to the edges of the supercell. Furthermore, it is clearly visible in Fig. \ref{fig:noccy0}(a) that for $V=0$ the charges are uniformly distributed with $\langle \hat{n}_{i(x,0)}\rangle\approx 2/3$. Again, as shown in Fig. \ref{fig:noccy0}(b), for $V=2.5|t|$ we observe a CDW modulation with the three triangular sublattices referring to $(\frac{1}{2}\frac{3}{4}\frac{3}{4})$. For $V = 4.0|t|$, the two sublattices are almost completely occupied and the remaining third sublattice has only residual occupancy [Fig.\ref{fig:noccy0}(c)]. Finally, for $V=6.0|t|$ the intersite repulsion overcomes the competitive Hubbard on-site interactions, and double occupancies are created on a single sublattice leaving the remaining two empty [Fig.\ref{fig:noccy0}(d)].

\begin{figure}
    \centering 
    \includegraphics[width=1.0\linewidth]{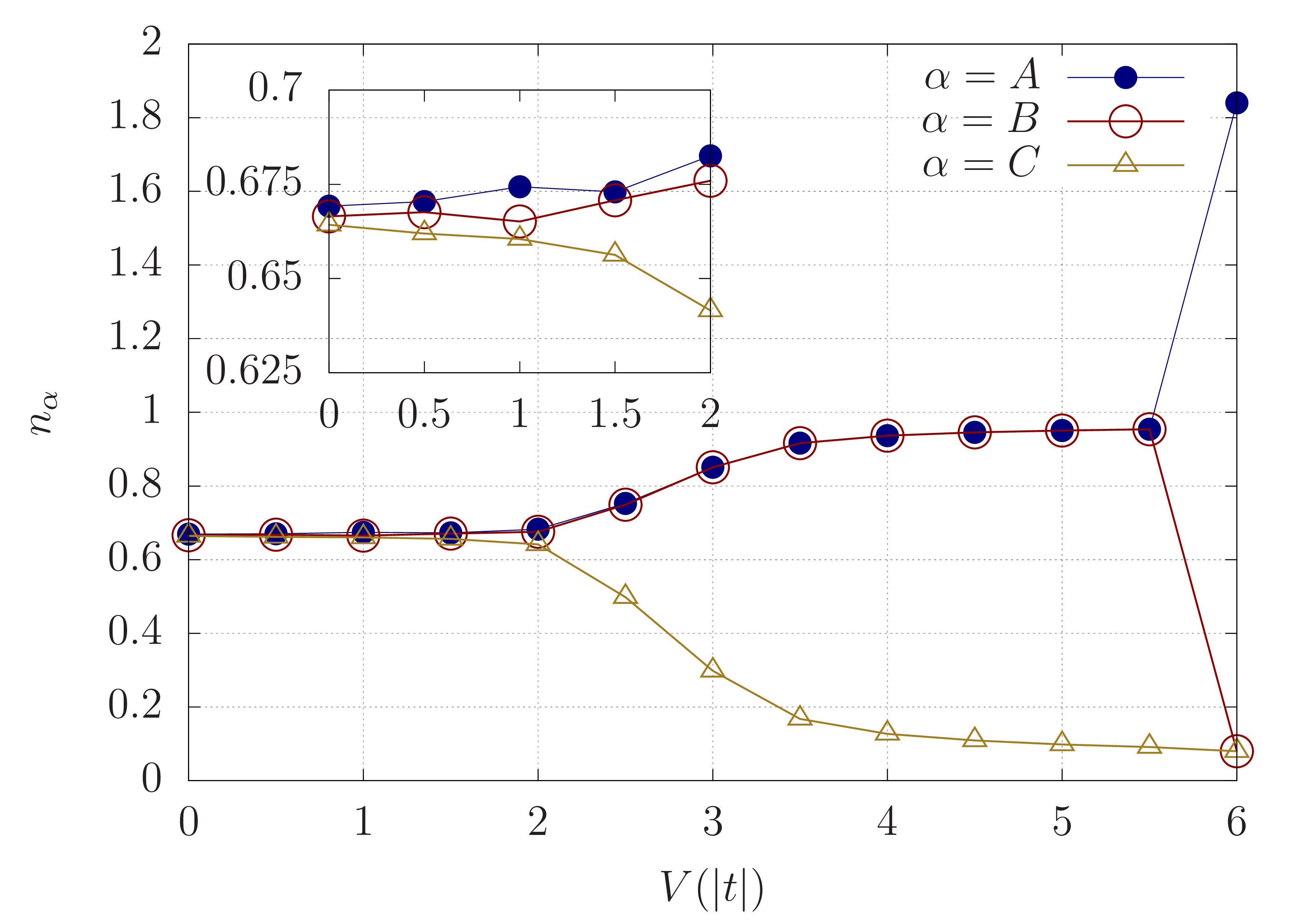}
    \caption{The electron occupation ${n}_{\alpha}$ (see the main text) of the three sublattices $A$, $B$, and $C$ as a function of intersite interaction $V$ for the system of size $24\times 6$.}
    \label{fig:cdw_order_summary}
\end{figure}

For the sake of completeness in Fig. \ref{fig:cdw_order_summary} we show the gradual evolution of the charge pattern induced by the presence of intersite Coulomb repulsion by plotting the average density $n_{\alpha}\equiv\frac{1}{3N}\sum_{i(\alpha)}\langle\hat{n}_{i(\alpha)}\rangle$ on each sublattice as a function of $V$. The differences between the three sublattices are rather weak for $V\lesssim1.5|t|$ (see the inset in Fig.\ref{fig:cdw_order_summary}). However, for $V\gtrsim1.5|t|$, a precoursor charge ordering appears and becomes more pronounced to finally achieve a nearly pure $(110)$ pattern for $V\approx4|t|$. This order is persistent up to $V\approx5.5|t|$, where an abrupt redistribution of charges leads to the emergence of the $(200)$ phase. Note that the emergence of the $(110)$ order takes place for $U/V\approx5$ and $(200)$ becomes stable for $U/V\approx3$. 

\begin{figure}
    \centering
    \includegraphics[width=0.7\linewidth]{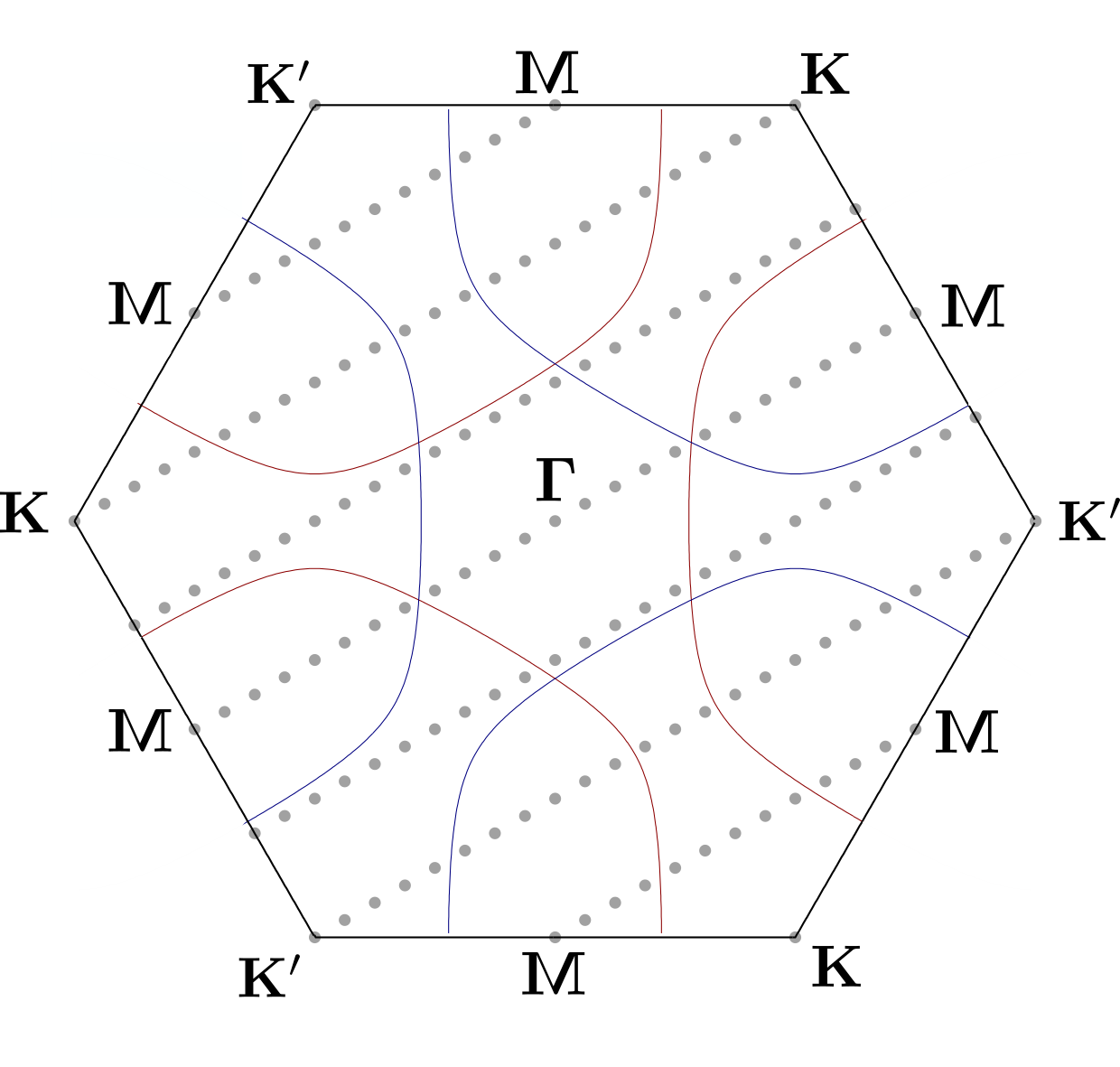}
    \caption{Fermi surfaces (solid lines) obtained by diagonalization of  the non-interacting part of Hamiltonian (i.e. for $\mathcal{\hat{H}}_0$)  at filling $n=2/3$. The solid red (blue) lines refer to the Fermi level of $\sigma=\uparrow(\downarrow)$ spin-valley split bands. The gray circles represent $\mathbf{q}$ vectors available for the finite  cylinder of size $24\times6$, that is the largest considered in the DMRG approach.}
    \label{fig:BZ}
\end{figure}

\begin{figure}
    \centering
    \includegraphics[width=1.0\linewidth]{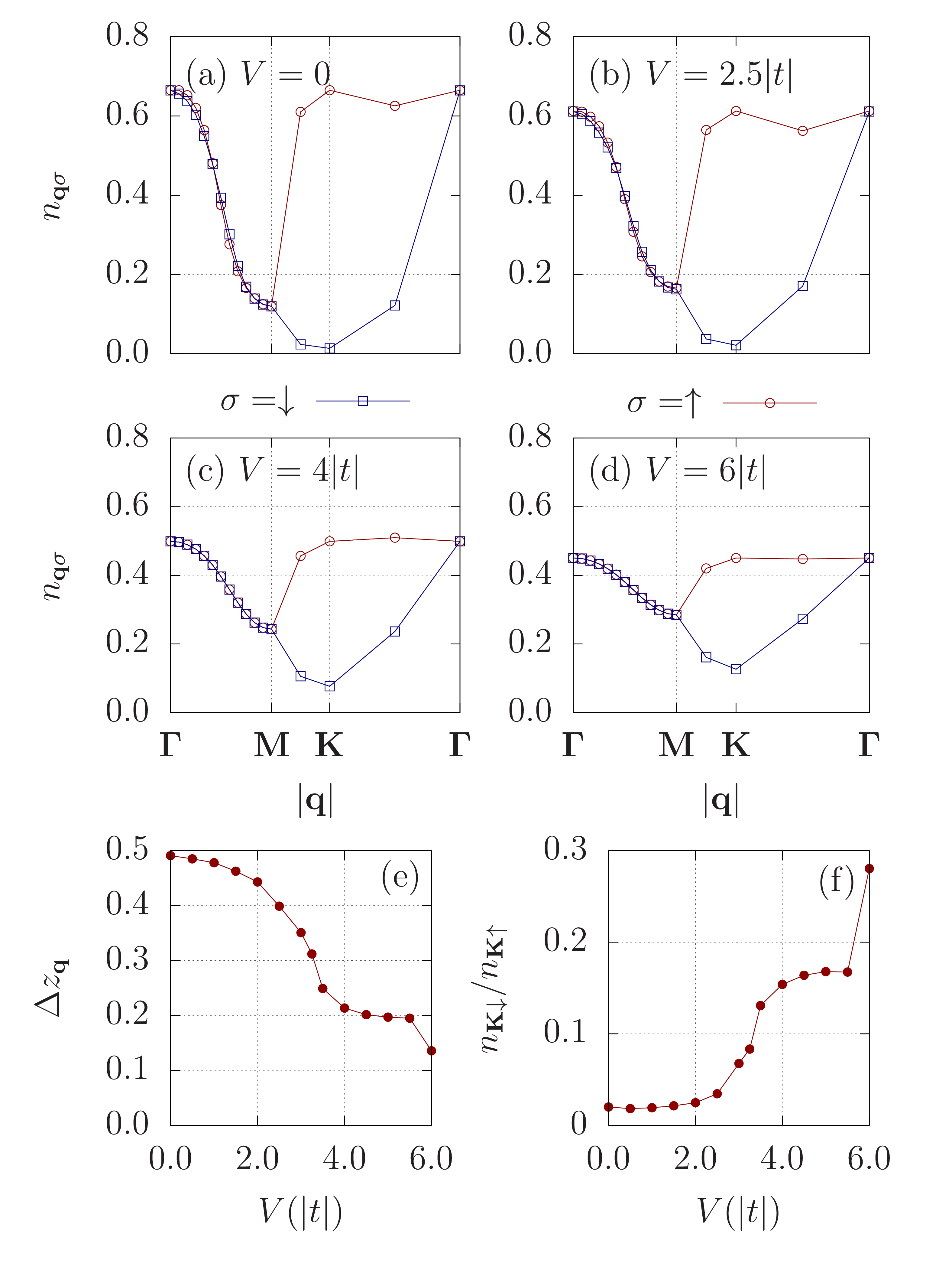}
    \caption{Momentum resolved occupation number as a function of $\mathbf{q}$ along the high symmetry directions in the reciprocal space for the four different values of $V$: (a) $V=0$; (b) $V=2.5|t|$; (c) $V=4.0|t|$ and (d) $V=6.0|t|$. Value of $\Delta z_\mathbf{q}$ (e), and the ratio  $n_{\mathbf{K}\downarrow}/n_{\mathbf{K}\uparrow}$ (f), both as a function of $V$.}
    \label{fig:nk}
\end{figure}

Subsequently, we investigate the momentum space-resolved single-particle correlation functions defined as
\begin{align}
    n_{\mathbf{q\sigma}}=\frac{1}{N}\sum_{i,j}\text{e}^{i\mathbf{q}\cdot(\mathbf{r}_i-\mathbf{r_j})}\langle\hat{c}_{i\sigma}^{\dagger}\hat{c}_{j\sigma}\rangle,
    \label{eq:nq}
\end{align}
where $\{\mathbf{r}_i\}$ are vectors pointing to the lattice sites and ${\mathbf{q}}$ are the momentum vectors. In Fig. \ref{fig:BZ} we present the available $\mathbf{q}$-space vectors for the cylinder of size $24\times6$, as well as the Fermi surfaces for both spin-split bands resulting from the diagonalization of $\mathcal{\hat{H}}_0$ in the reciprocal space at $n=2/3$. As one can see, in the non-interacting picture, Fermi surfaces are opened around $\mathbf{K}$ and $\mathbf{K}^{\prime}$ points for $\delta=\uparrow$ and $\delta=\downarrow$ quasiparticles, respectively. The presence of $C_3$ symmetry for both spin subbands separately is an expression of the SOC encapsulated in $\mathcal{\hat{H}}_0$ that breaks the $C_6$ symmetry of the triangular lattice and results in the appearance of spin-valley locking~\cite{Tan2023}. 
The analysis of $n_{\mathbf{q}\sigma}$ may provide useful information with respect to the quasiparticles \emph{dressed} in interactions ~\cite{Spalek2022}. That is, the quasiparticle weight $Z_\mathbf{q}$ that measures the coherence in the fermionic interacting system~\cite{Spalek2022} may serve as an indicator of the electron correlation strength. In Figs. \ref{fig:nk}(a-d) we show $n_{\mathbf{q}\sigma}$ along the $\Gamma-M-K-\Gamma$ path for the four representative values of $V$. Note that along the $\Gamma-M-K^{\prime}-\Gamma$ trajectory the result is identical when one exchanges spin-up with spin-down quasiparticles. As one can see in Fig. \ref{fig:nk}(a), even for $V=0$, the quasiparticles are renormalized due to the high value of $U$, since we do not observe an abrupt decrease in $n_\mathbf{q\sigma}$ for $\mathbf{q}\in \mathbf{M}-\mathbf{\Gamma}$ in the vicinity of the point where the two branches of Fermi surfaces cross (see Fig.\ref{fig:BZ}). The similar observation holds for $V=2.5|t|,4.0|t|$ and $6.0|t|$ as shown in Figs. \ref{fig:nk}(b-d). 
Although the $\mathbf{K}-\mathbf{M}$ section is affected by a lower density of $\mathbf{q}$ points, an abrupt decrease in $n_{\mathbf{q}\uparrow}$($n_{\mathbf{q}\downarrow}$) seems to appear at $\mathbf{q}_{(K+M)/2}\equiv{(\mathbf{K}+\mathbf{M})/2}$. Apparently, at $\mathbf{q}_{(K+M)/2}$, the occupancy, $n_{\mathbf{q}}$, is higher by a value of $\approx0.4$ compared to $\mathbf{q}=\mathbf{M}$. For $\sigma=\downarrow$, it concomitantly decreases to $\sim0.0$ for $V=0$, while for the higher values of intersite repulsions, the corresponding values become higher with increasing $V$. This, together with the general \emph{flattening} of $n_{\mathbf{q}\sigma}$, signals further renormalization of quasiparticles driven by the increase of $V$. For $\mathbf{q}\in\mathbf{\Gamma}-\mathbf{K}$, it can be deduced from Figs. \ref{fig:nk}(a-d) that $n_{\mathbf{q}\uparrow}$ remains nearly constant, while $n_{\mathbf{q}\downarrow}$ rapidly increases when the Fermi surface is crossed. Subsequently, the latter attains the same value as for spin-up carriers, as expected.

The vanishing of the quasiparticle weight  $Z_{\mathbf{q}}$ can be regarded as an indicator of the insulating character of the interacting system \cite{Brinkman1970,Senthil,Spalek2022}. As we consider a supercell of finite size and it is the numerically demanding task to obtain the precise value of the $Z_{\mathbf{q}}$ we are unable undoubtedly to deduce if interactions drive the system to the insulating state in this manner. However to gain some intuition in this regard we inspect quantity 
\begin{align}
\Delta z_{\mathbf{q}}\approx n_{\mathbf{q}_{(K+M)/2}\uparrow}-n_{\mathbf{M}\uparrow}=n_{\mathbf{q}_{(K'+M)/2}\downarrow}-n_{\mathbf{M}\downarrow}
\label{eq:deltaz}
\end{align}
as a function of $V$ instead, as it provides some information in view of the quasiparticle renormalization. Consequently, from Fig.\ref{fig:nk}e one deduces that for $V\lesssim3|t|$, the intersite interactions  sites moderately modify $n_{\mathbf{q}\sigma}$ compared to the case $V=0$. When approaching $V\approx3|t|$, the renormalization is significantly enhanced. For $3.5|t|\lesssim V  \lesssim5.5|t|$ no spectacular change can be observed in $\Delta z_{\mathbf{q}}$, however, for $V=6|t|$ an abrupt decrease associated with the formation of the $(200)$ ordered phase appears, indicating an extreme reorganization of the occupation scheme in the momentum space. The role of inter-site repulsion on spin-valley polarization can also be identified by inspecting the ratio $n_{\mathbf{K}\uparrow}/n_{\mathbf{K}\downarrow}$ (or alternatively $n_{\mathbf{K}'\downarrow}/n_{\mathbf{K}'\uparrow}$) as presented in Fig.\ref{fig:nk}f. That is, interactions-driven \emph{smearing} of Fermi surfaces result in decreased $n_{\mathbf{K}\uparrow}$ accompanied by increased occupation of $\mathbf{K}$ momentum state by $\sigma=\downarrow$ quasiparticles. This behavior illustrates the reduction of anisotropy with respect to spin quantum number along the path $\mathbf{M}-\mathbf{K}-\mathbf{\Gamma}$ in the single-particle occupation scheme with an increasing value of $V$.

To further characterize the electronic properties of the system in view of possible insulating gap opening due to long-range Coulomb repulsion and in turn the formation of the GWC state, we investigate the Fourier transform of the two-body density-density correlation functions expressed as
\begin{align}
    \mathcal{N}(\mathbf{q})=\frac{1}{N}\sum_{ij}\text{e}^{i\mathbf{q\cdot(\mathbf{r}_i-\mathbf{r}_j)}}\langle \hat{n}_i\hat{n}_j\rangle.
\end{align}
That is, $\lim_{|\mathbf{q}|\rightarrow0}\mathbf{q}^2/\mathcal{N}(\mathbf{q})$ is proportional to the magnitude of the gap~\cite{Becca2005,Tocchio2011,Tocchio2014,Tocchio2020,Biborski2024}. According to the finite size of the system, it is convenient to inspect $\mathcal{N}(\mathbf{q})/|\mathbf{q}|$ for the smallest available value of $|\mathbf{q}|$. In fact, when $\mathcal{N}(\mathbf{q})\sim |\mathbf{q}|^2$ for $|\mathbf{q}|\rightarrow0$ it indicates the opening of the gap~\cite{Tocchio2020}. Although our approach breaks the translational symmetry due to the application of open boundary conditions in one direction, we find that the length of the cylinder considered here (i.e. $L_1=24$) is large enough to observe, at least qualitatively, an abrupt change in the character of $\mathcal{N}(\mathbf{q})/|\mathbf{q}|$ with increasing value of $V$.

\begin{figure}
    \centering
    \includegraphics[width=1.0\linewidth]{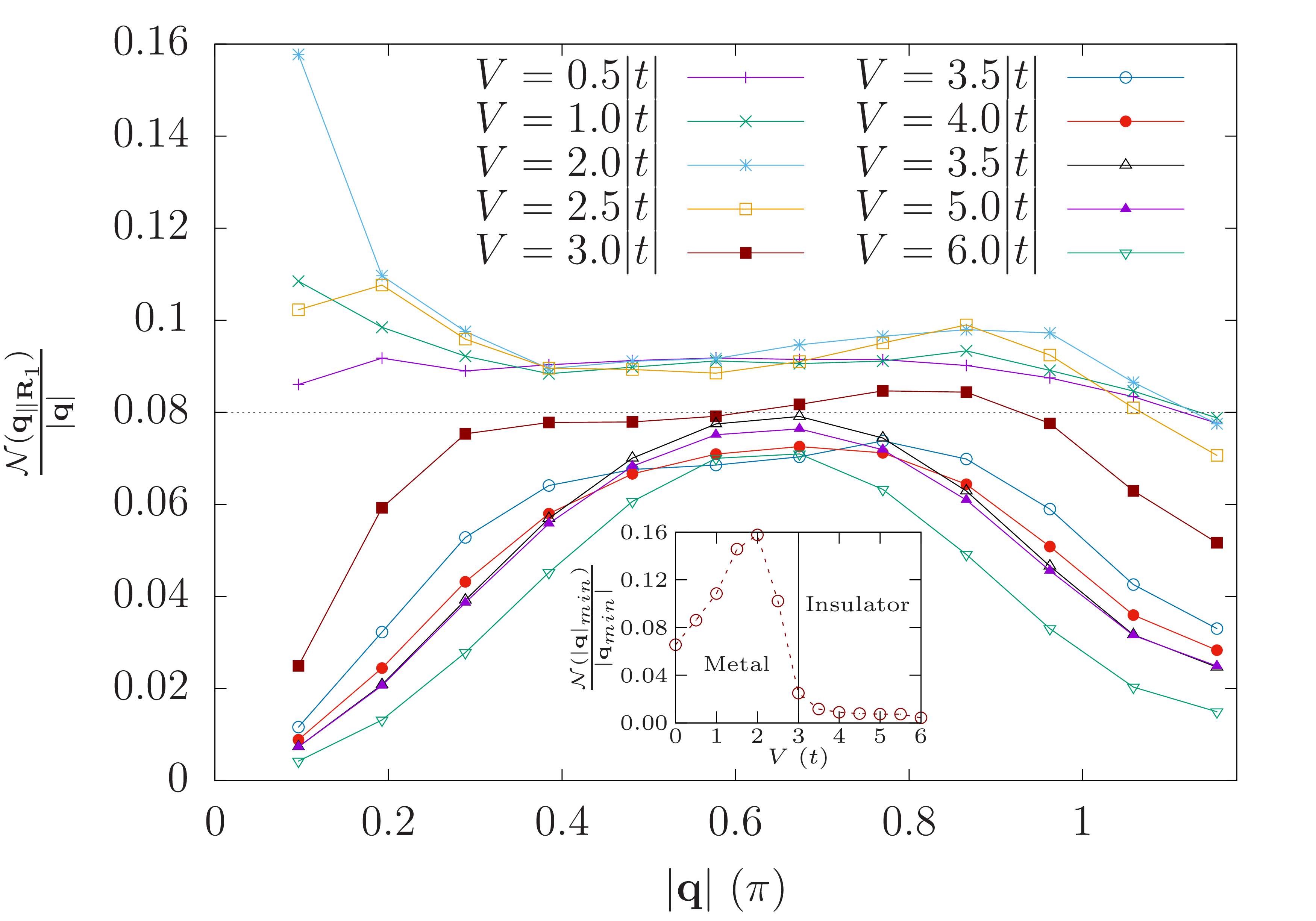}
    \caption{The $\mathcal{N}(\mathbf{q})/|\mathbf{q}|$  dependence for selected values of $V$. The horizontal line at $0.08$ roughly divides the curves referring to the metallic state from those showing insulating behavior (although the case $V = 0$ is an isolated exception to this rule). The inset shows the same quantity as the Figure itself, but for the smallest available $|\mathbf{q}|$ as function of $V$.}
    \label{fig:nnq}
\end{figure}

In Fig. \ref{fig:nnq}, we plot this quantity for $\mathbf{q}$ along $\mathbf{\Gamma}-\mathbf{M}$. Although the dependencies obtained for $V\lesssim3|t|$ in the vicinity of $|\mathbf{q}|\rightarrow 0$ do not expose a regular pattern with increasing value of $V$ they are clearly distinguishable from those obtained for $V\gtrsim3|t|$. That is, for the latter case the values of $\mathcal{N}(\mathbf{q})/|\mathbf{q}|$ clearly decrease when $|\mathbf{q}|$ is small. This trend can also be seen in the inset of Fig. \ref{fig:nnq} where we plot the same quantity as in the main figure (as a function of $V$) but only for the smallest value of $|\mathbf{q}|$ available in our calculations. 
In this view, just below $V\approx3|t|\sim 3.5|t|$ the system exhibits a rather extreme form of the CDW pattern, since the formation of GWC is supposed to be concomitant with the opening of the gap~\cite{Hubbard1978,Padhi2018,Regan2020} induced by the intersite Coulomb repulsions.   Therefore, based on the above reasoning and the inspection of the magnetic properties of the system provided in the following, we estimate that the threshold for the emergence of the GWC state is $V\approx3|t|\sim 3.5|t|$.  We find this value reasonable, since the approach based on the variational Monte Carlo method for the extended Hubbard model on the triangular lattice \cite{Watanabe2005} provides critical $V\approx2.7|t|$ at $U=14|t|$. Also, as comes from the $(U,V)$ phase diagram for $n=2/3$ presented by Watanabe and Ogata \cite{Watanabe2005} one may expect that critical $V$ may slightly decrease when one increases $U$.


\subsection{Spin order}
The analysis of the charge distribution carried out in the previous subsection allows us to state that the considered model reconstructs the experimental situation from the point of view of the GWC appearance in the WSe$_2$/WS$_2$ heterobilayer. In this subsection, we characterize the spin properties of the system, focusing on the $(110)$ GWC case. 

\subsubsection{Correlation functions in the real space}
\begin{figure}
    \centering
    \includegraphics[width=1.0\linewidth]{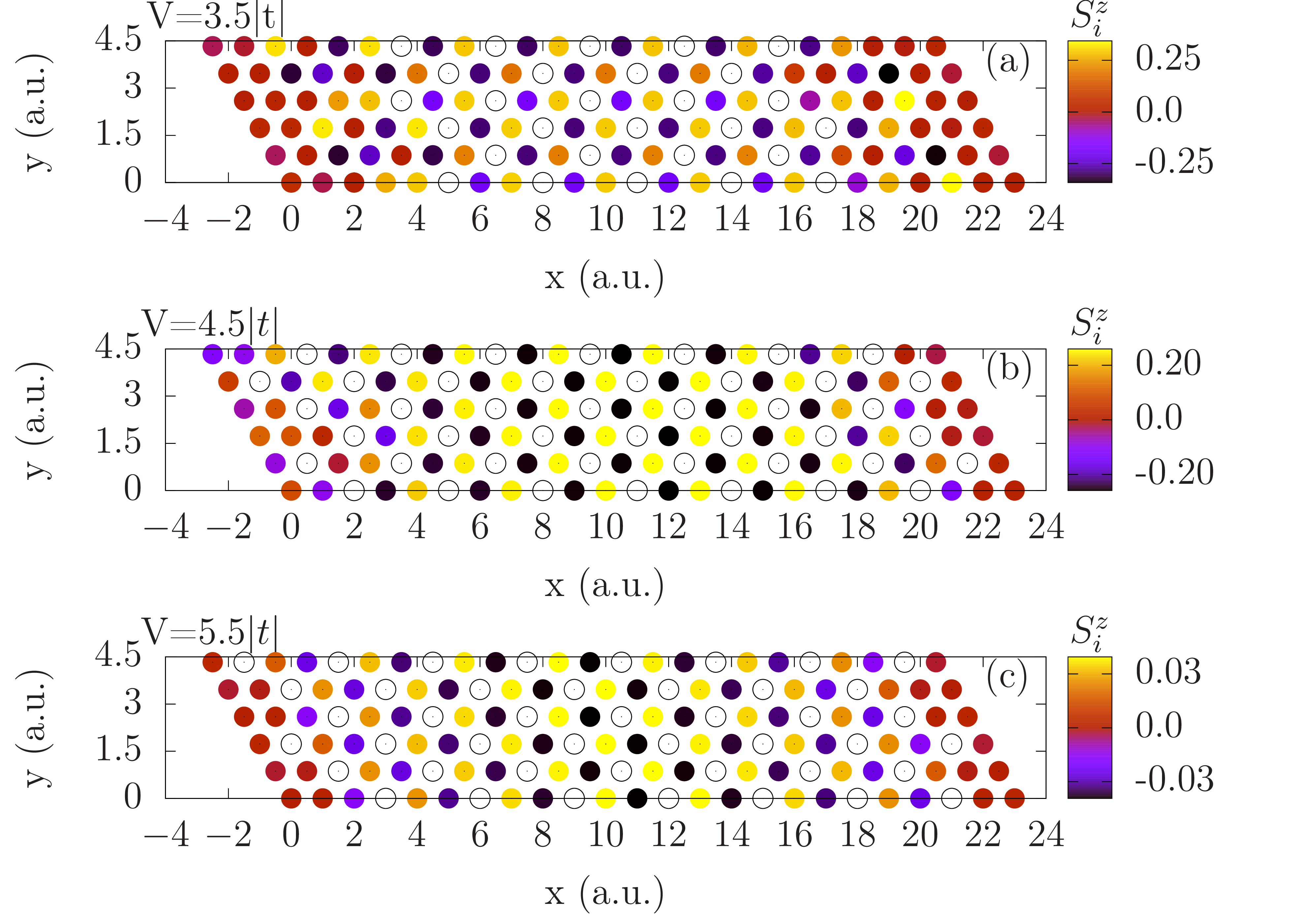}
    \caption{The real space resolved $z$-component of spin for the three representative values of $V$, that is: $V=3.5|t|$ (a), $V=4.5|t|$ (b), and $V=5.5|t|$ (c). Empty circles indicates sites where the mean occupation by carriers is lower than 0.1 threshold.}
    \label{fig:szonsite}
\end{figure}

The Hamiltonian considered here is not SU(2) invariant~\cite{Zang2021} since $\phi\neq n\pi$. Therefore, we focus on analysis of the in- and out-of-plane spin correlation functions separately. 
As first we inspect mean value of z-component of spin defined in the standard manner, i.e., as
\begin{align}
{S}_{i}^{z}=\Big\langle\frac{1}{2}(\hat{n}_{i\uparrow}-\hat{n}_{i\downarrow})\Big\rangle. 
\label{eq:sz}
\end{align}
The DMRG algorithm spontaneously converges towards  non-zero $S_{i}^z$ as can be deduced from Fig.\ref{fig:szonsite} where we plot $S^z_i$ for the three representative values of $V$  referring to $(110)$ charge order. Unlike edge effects that are more pronounced for the lower values of $V$, the staggered pattern of spins is clearly manifested. However, we observe that the absolute magnitude of $S_i^z$ decreases as the amplitude of $V$ increases. 

To gain insight into the character of magnetic ordering, we subsequently analyze both the out-of-plane and in-plane components of the spin-spin correlation functions. Since  the resultant states are biased by a non-zero $S_i^{z}$ we study the out-of-plane part in the form 
\begin{align}
\mathcal{S}_{ij}^{z}=\Big\langle\hat{S}_i^{z}\hat{S}_j^z\Big\rangle - \langle \hat{S}_i^{z}\rangle\langle \hat{S}_j^{z}\rangle.
\label{eq:szsz}
\end{align}
We also inspect in-plane component 
\begin{align}
\mathcal{S}_{ij}^{xx}=\mathcal{S}_{ij}^{yy}=\Big\langle\frac{\hat{S}_i^{+}\hat{S}_j^{-}+\hat{S}_i^{-}\hat{S}_j^{+}}{4}\Big\rangle,
\label{eq:sxsx}
\end{align}
where $\hat{S}^{+}$ ($\hat{S}^{-}$) is spin rising(lowering) operator. Note the equality $\mathcal{S}_{ij}^{xx}=\mathcal{S}_{ij}^{yy}$ that is valid since we work in the $S_{tot}^{z}=0$ sector. However, since the Hamiltonian considered here is not \emph{explicitly} SU(2) invariant (though it has \emph{hidden} SU(2) invariance ~\cite{Zang2021,Motruk2023} as will be discussed later on), the correlations between $x$ and $y$ components of spin, i.e.,
\begin{align}
\mathcal{S}_{ij}^{xy}=-\mathcal{S}_{ij}^{yx}=\Big\langle\frac{\hat{S}_i^{-}\hat{S}_j^{+}-\hat{S}_i^{+}\hat{S}_j^{-}}{4i}\Big\rangle,
\label{eq:sxsy}
\end{align}
possibly do not vanish, Consequently, we inspect $\mathcal{S}_{ij}^{z}$, $\mathcal{S}_{ij}^{xx}$, and $\mathcal{S}_{ij}^{xy}$ separately to elucidate the spin ordering basing on the complete set of correlations.

\begin{figure}
    \centering
    \includegraphics[width=1.0\linewidth]{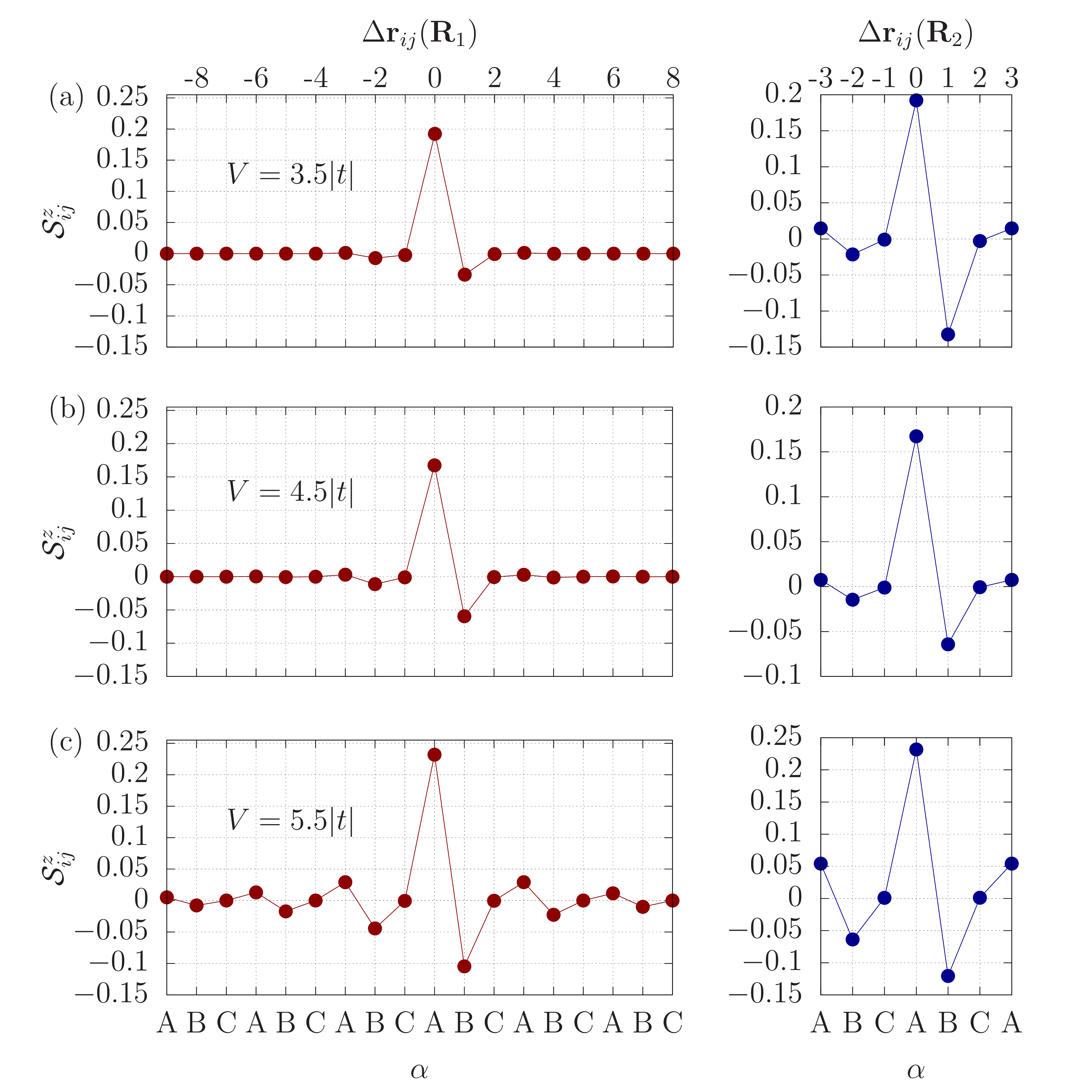}
    \caption{The correlation functions corresponding to the $z$ component of spin for the three representative values of $V$, that is: $V=3.5|t|$ (a), $V=4.5|t|$ (b), and $V=5.5|t|$. The left panel correspond to correlation functions collected along $\mathbf{R}_1$ direction whereas the right one to those along $\mathbf{R}_2$.}
    \label{fig:szsz}
\end{figure}

In Fig. \ref{fig:szsz}, we present the spatial dependence of $\mathcal{S}_{ij}^{z}$. We observe the presence of fast decaying antiferromagnetic correlations up to $V\approx5|t|$, however, the sign clearly oscillates indicating a tendency towards antiferromagnetic order. As can be seen for $V\gtrsim5|t|$ the absolute value of $\mathcal{S}_{ij}^{z}$ becomes characterized by a weaker spatial decay. Thus, at least in view of the $z$ -th component of spin, the AF order becomes more robust with increasing $V$.

\begin{figure}
    \centering
    \includegraphics[width=1.0\linewidth]{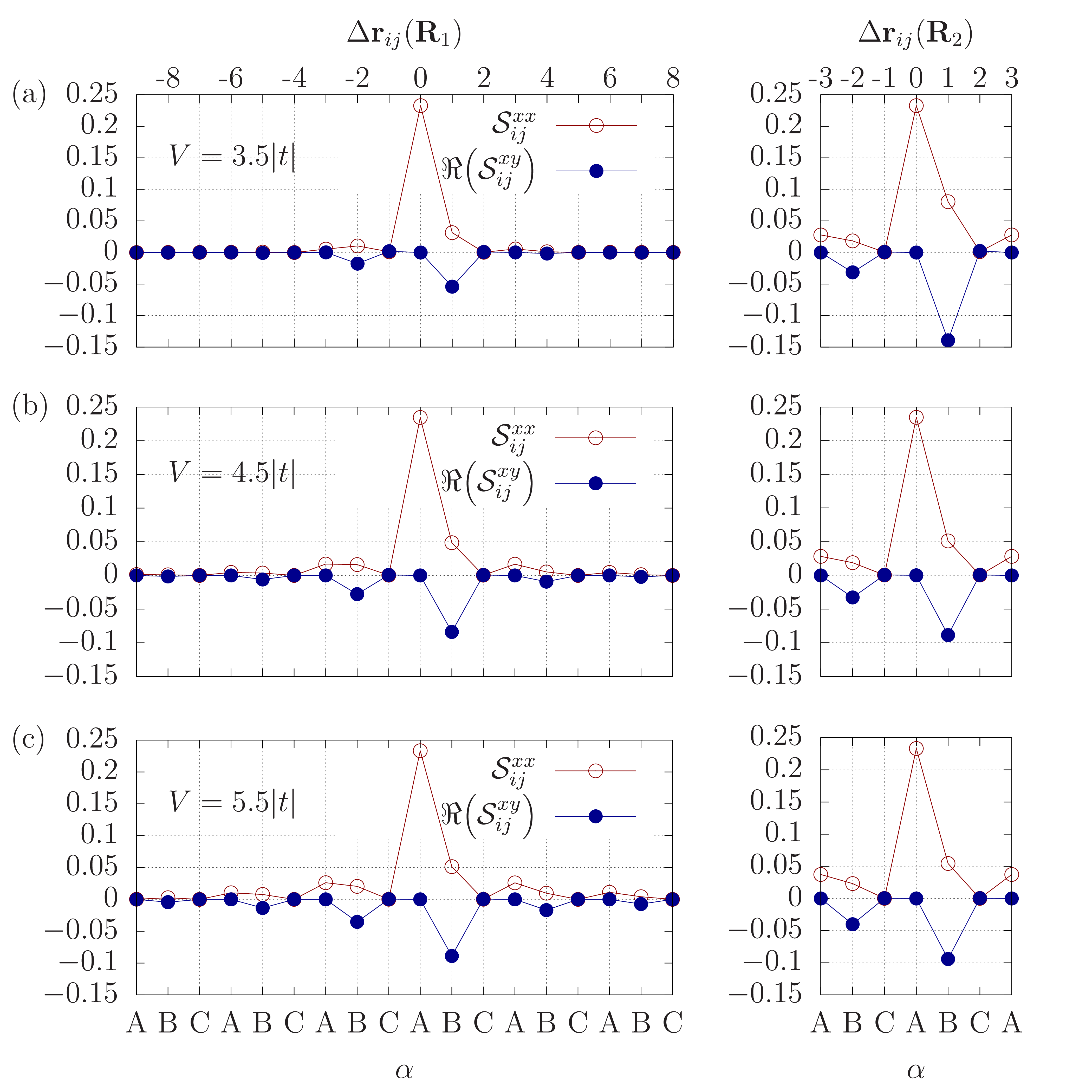}
    \caption{The spin-spin correlation function $S_{ij}^{xx}$ and the real parts of $S_{ij}^{xy}$  for the selected values of $V$, that is: $V=3.5|t|$ (a), $V=4.5|t|$ (b), and $V=5.5|t|$ (c). The left panel correspond to correlation functions collected along $\mathbf{R}_1$ direction whereas the right one to those along $\mathbf{R}_2$. The increase in both the magnitude and spatial extension is clearly visible when $V$ approaches range for which system becomes close to $(200)$ phase.}
    \label{fig:spsm}
\end{figure}

\begin{figure}
    \centering
    \includegraphics[width=1.0\linewidth]{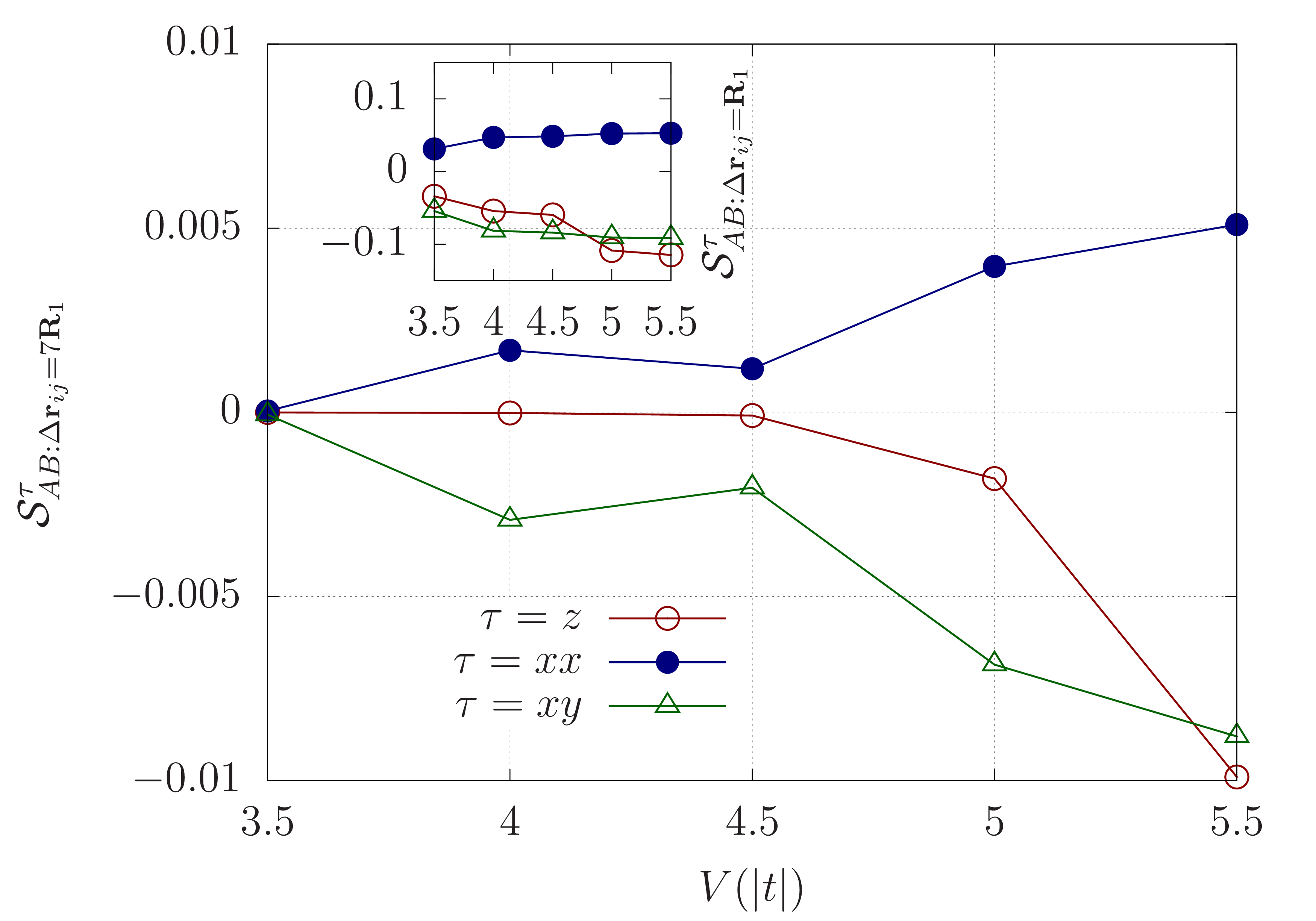}
    \caption{The resultant $\mathcal{S}_{AB}^{z}$, $\mathcal{S}_{AB}^{xx}$ and $\mathcal{S}_{AB}^{xy}$ (which is purely real) for $(110)$ phase as a function of $V$ collected for $\Delta{\mathbf r}_{ij}=7\Delta \mathbf{R}_1$ (main plot) and for $\Delta{\mathbf r}_{ij}=\Delta \mathbf{R}_1$ (inset).}
    \label{fig:szsxy}
\end{figure}

The enhancement in magnetic correlations with increasing magnitude of $V$ is expressed also in terms of the correlation functions of $S^{xx}_{ij}$ and $S^{xy}_{ij}$ as shown in Fig.\ref{fig:spsm}. The correlations $S^{xx}_{ij} (S^{yy}_{ij})$ are solely positive between occupied sites indicating a ferromagnetic pattern, which, however, is \emph{canted} since we find $arg\Big(\langle \hat{S}_{i}^{+}\hat{S}_{j}^{-}\rangle\Big)\approx\pm\frac{\pi}{3}$. We also observe the presence of a non-zero real part of the $S_{ij}^{xy}$ correlations for $i\neq j$  (see Fig.\ref{fig:spsm}). This can be interpreted as the emanation of breaking of SU(2) symmetry in the considered Hamiltonian. Note that we find the imaginary part of $S_{ij}^{xy}$ non-zero only when $i=j$ since we have $S_{i}^{z}\neq0$ in most cases when $i\in\{A,B\}$.
However, the spatial decay of the in-plane spin-spin correlations has a similar characteristic to those in the $z$ direction; that is, they decay noticeably faster when $V\lesssim5|t|$.
We conclude that spin-spin correlations are enhanced when $V$ increases. Our findings based on the analysis carried out in the real space are compactly presented in Fig.\ref{fig:szsxy} where in the main plot, we consider the case where the sublattice sites $A$ and $B$ are separated by $\Delta{\mathbf r}_{ij}=7\Delta \mathbf{R}_1$, that is, the range for which the result is still not affected by open boundary conditions. Eventually, the vital magnetic order is supposed to emerge in the effective honeycomb lattice when the residual charge occupancy on the sublattice $C$ decreases further. 

\subsubsection{Spin order and gauge transformation}
As mentioned above, the existence of $S^{xy}_{ij}\neq0$ can serve as an indicator of breaking of SU(2) invariance. Thus, one may consider finding a single particle basis $\{\hat{d}_{i,\sigma}\}$ in which the correlations $\Tilde{S}^{xy}_{ij}$ given in this new basis vanish. For the particular case studied here, that is, for $\phi=\frac{2\pi}{3}$ such a procedure is possible. As provided by Zang et al. ~\cite{Zang2021} there exists the \emph{gauge} transformation $\phi\rightarrow\phi\pm\frac{2\pi}{3}$ which does not change the energetic spectrum of the Hamiltonian albeit modifies the corresponding eigenstates by some unitary transformation $\hat{\Phi}_{\frac{2\pi}{3}}^{\dagger}=\hat{\Phi}_{-\frac{2\pi}{3}}=\hat{\Phi}_{\frac{2\pi}{3}}^{-1}$ which maps $\{\hat{a}_{i,\sigma}\}$ onto $\{\hat{d}_{i,\sigma}\}$. Since in the model analyzed by us $\phi=\frac{2\pi}{3}$ it is possible to verify its spin properties by applying this unitary transformation to the resulting ground state and compare them with the solution obtained directly by solving the SU(2) invariant Hamiltonian for which $\phi=0$. A similar procedure has recently been performed for filling $n=3/4$ referring to the formation of kagome GWC ~\cite{Motruk2023}. From the spin-spin correlation perspective, this procedure can be considered as finding the basis on which the correlations described by $S^{xy}_{ij}$ are absent, as will be shown.
We believe that the details of this unitary transformation have been thoroughly studied~\cite{Zang2021,Zang2022,Motruk2023}, however, for the sake of clarity, we sketch it below.

In the first step $A,B$ and $C$ the lattice sites are grouped into triangular plaquettes that are now indexed by $p$ and each site is further labeled by the index $\alpha\in\{A,B,C\}$. Next, operators $\hat{a}_{\alpha,p,\sigma}$ are transformed applying  the following recipe which delivers $\hat{\Phi}_{\frac{2\pi}{3}}$:
\begin{align}
\begin{split}\label{eq:utransfrom}
    \hat{a}_{A,p,\sigma} \rightarrow{}& \hat{a}_{A,p,\sigma}\equiv\hat{d}_{A,p,\sigma}\\
    \hat{a}_{B,p,\sigma} \rightarrow& \text{e}^{-i\sigma\frac{2\pi}{3}}\hat{a}_{B,p,\sigma}\equiv\hat{d}_{B,p,\sigma}\\
    \hat{a}_{C,p,\sigma} \rightarrow{}& \text{e}^{i\sigma\frac{2\pi}{3}}\hat{a}_{C,p,\sigma}\equiv \hat{d}_{C,p,\sigma}.
\end{split}
\end{align}
The operators $\hat{d}_{\alpha,p,\sigma}$ form the new single-particle basis and also conform to standard fermionic anti-commutation relations. 
Let us assign the ground state of the SU(2) invariant Hamiltonian $\mathcal{\hat{H}}_{SU(2)}$ (that is, with $\phi=0$) as $|\Psi_{SU(2)}\rangle$. Then the ground state of the Hamiltonian considered here, that is, of $\mathcal{\hat{H}}=\hat{\Phi}_{\frac{2\pi}{3}}\mathcal{\hat{H}}_{SU(2)}\hat{\Phi}_{\frac{2\pi}{3}}^{\dagger}$ is $|\Psi\rangle=\Phi_{\frac{2\pi}{3}}|\Psi_{SU(2)}\rangle$ since $\hat{\Phi}_{\frac{2\pi}{3}}^{\dagger}{\Phi}_{\frac{2\pi}{3}}=\hat{I}$. The energy spectrum remains unchanged because $\langle\Psi_{SU(2)}|\mathcal{\hat{H}}_{SU(2)}|\Psi_{SU(2)}\rangle=\langle\Psi|\hat{\Phi}_{\frac{2\pi}{3}}^{\dagger}\hat{\Phi}_{\frac{2\pi}{3}}\mathcal{\hat{H}}\hat{\Phi}_{\frac{2\pi}{3}}^{\dagger}\hat{\Phi}_{\frac{2\pi}{3}}|\Psi\rangle$.

In the language of spin operators, this transformation results in the sublattice-resolved spin rotation in the $x-y$ plane~\cite{Zang2021}. Namely, one finds that spin operators $\hat{\Tilde{S}}_{\alpha,p}^{x/y}$ are now related with $\hat{{S}}_{\alpha,p}^{x/y}$ by
\begin{align}
\mathbf{\hat{s}}_{\alpha,p}\equiv
\begin{pmatrix}
\hat{\Tilde{S}}^{x}_{\alpha,p}\\
\hat{\Tilde{S}}^{y}_{\alpha,p}
\end{pmatrix}=
\mathbf{R}^{\varphi_\alpha}
    \begin{pmatrix}
\hat{{S}}^{x}_{\alpha,p}\\
\hat{{S}}^{y}_{\alpha,p}
\end{pmatrix},
\label{eq:spinrot}
\end{align}
where $\mathbf{R}^{\varphi_\alpha}$ is  matrix providing rotations  by the angle selected depending on $\alpha$, explicitly it is $\varphi_A=0$, $\varphi_B=\frac{2\pi}{3}$ and $\varphi_C=-\frac{2\pi}{3}$.  According to the relations given above, the operators utilized for computing correlation function after performing this transformation  are in a compact notation given as
\begin{align}
\hat{\tilde{S}}^{\tau}_{\alpha,p}\hat{\tilde{S}}^{\tau'}_{\alpha',q}=\Big[\mathbf{\hat{s}}_{\alpha,p}^\mathbf{T}\otimes\mathbf{\hat{s}}_{\alpha',q}\Big]_{\tau\tau'}=&\nonumber\\=\sum_{ij}\hat{S}^{i}_{\alpha,p}\hat{S}^{j}_{\alpha',q}(\mathbf{R}^{\varphi_{\alpha}})^{\mathbf{T}}_{i\tau}\mathbf{R}^{\varphi_{\alpha'}}_{\tau'j},
\label{eq:xxtrans}
\end{align}
where $\tau\in\{x,y\}$ (the same holds for $i,j$ indices).  
The $\hat{\Phi}_{\frac{-2\pi}{3}}$ by construction transforms the Hamiltonian with $\phi=2\pi/3$ into the SU(2) invariant form. However it is still instructive  to consider also the real parts of correlation functions associated with those operators in Eq.\ref{eq:xxtrans} for which $\tau\neq \tau'$, e.g.,
\begin{align}
\hat{\Tilde{S}}^{x}_{A,p}\hat{\Tilde{S}}^{y}_{B/C,q}=\pm\sin{\frac{2\pi}{3}}\Big({\hat{S}^{x}_{A,p}\hat{S}^{x}_{B/C,q}}\Big)+\nonumber\\+\cos{\frac{2\pi}{3}}\Big({\hat{S}^{x}_{A,p}\hat{S}^{y}_{B/C,q}}\Big),
\label{eq:saxsybc}
\end{align}
since they should vanish  and can be juxtaposed with the numerical data obtained directly for $\phi=0$ as mentioned above. 

\begin{figure}
    \centering
    \includegraphics[width=1.0\linewidth]{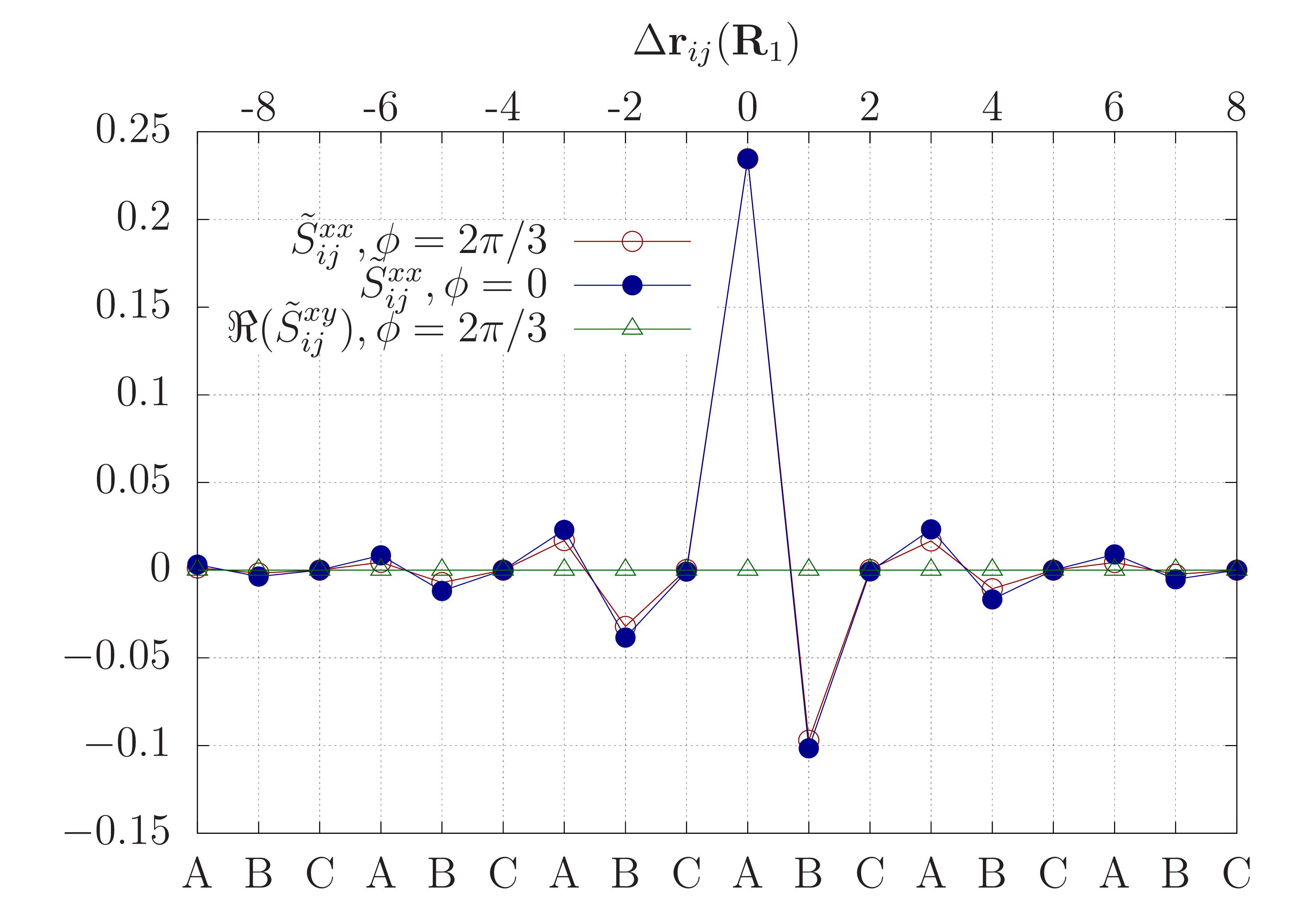}
    \caption{Transformed correlation function $\Tilde{\mathcal{S}}^{xx}_{ij}$ and
    the real part of $\Tilde{\mathcal{S}}^{xy}_{ij}$ both obtained  at $V=4.5|t|$ and for $\phi=\frac{2\pi}{3}$. As can be deduced $\Tilde{\mathcal{S}}^{xx}_{ij}$ exhibits nearly perfect agreement with the non-transformed correlation function ${S}^{xx}_{ij}$ resulting from the calculations in which $\phi=0$.}
    \label{fig:sxxsyytrans}
\end{figure}

In Fig.\ref{fig:sxxsyytrans} we present correlation functions $\Tilde{\mathcal{S}}^{xx}_{ij}$ and $\Re({\Tilde{\mathcal{S}}^{xy}_{ij}})$ obtained using Eq.\ref{eq:xxtrans} for the case when $V=4.5|t|$. We observe that the $\Tilde{\mathcal{S}}^{xy}_{ij}$ correlations vanish as predicted. The $\Tilde{\mathcal{S}}^{xx}_{ij}$ correlations indicate antiferromagnetic order now. We subsequently validated this result, confronting the latter with $\mathcal{S}^{xx}_{ij}$ resulting from the auxiliary calculations in which we explicitly assumed $\phi=0$ in the Hamiltonian. We have found nearly excellent agreement between both approaches as can be deduced from Fig.\ref{fig:sxxsyytrans}. 
Therefore, we find the approach based on the transformation given in Eq.\ref{eq:utransfrom} useful in view of the validation of the numerical procedure. However, the experimental scenario which we address in this study concerns a system in which the SU(2) symmetry is broken. This results in the specific form of the complex-valued hopping terms in the effective model of WSe$_2$/WS$_{2}$  heterobilayer. It is worth highlighting in this context that in the WSe$_{2}$/WSe$_2$ homobilayers where $\phi$ is tunable in terms of the external electric field (displacement field),  the possibility of emulation of the SU(2) invariant Heisenberg model is disscussed~\cite{Kennes2022}.

\begin{figure}
    \centering
    \includegraphics[width=0.99\linewidth]{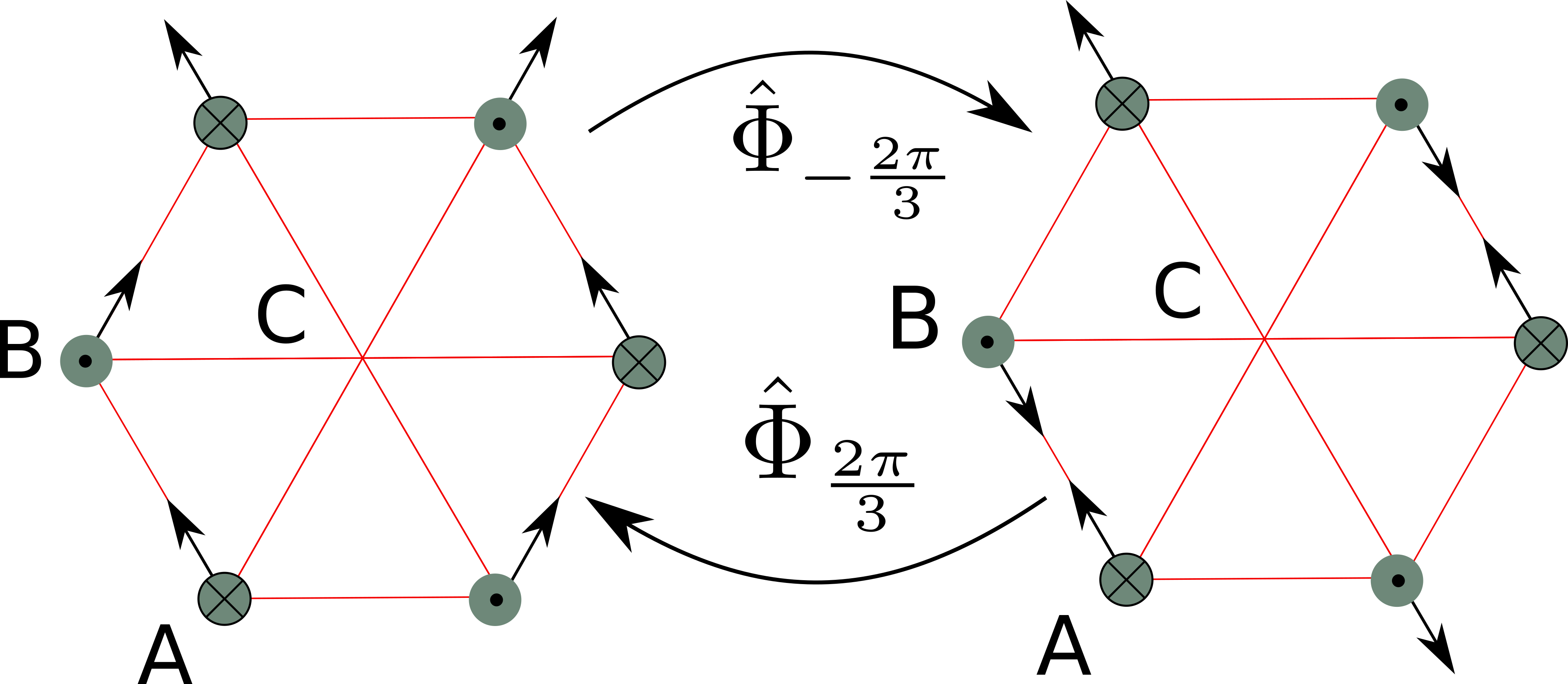}
    \caption{The sketch of the identified spin order in the GWC phase under consideration. Note, that for $\phi=\frac{2\pi}{3}$ (hexagon on the left) we observe \emph{canted} ferromagnetic order whereas the out-of-plane component (green circles) is antiferromagnatic. Eventually resulting state is also canted. For $\phi=0$, one obtains regular AF  pattern (hexagon on the right). One may obtain the canted spin order also by applying $\hat{\Phi}_{\frac{2\pi}{3}}$ to the ground state resulting from the diagonalization of SU(2) invariant Hamiltonian.}
    \label{fig:sketch}
\end{figure}

Recapitulating, by inspecting spin-spin correlation functions in the real-space picture, we reveal the tendency toward \emph{canted} ferromagnetic ordering in the $x-y$ plane, which is associated with SU(2) symmetry breaking in the Hamiltonian. However, the $z$-th component of antiferromagnetic spin-spin correlations is present for both $\phi=\frac{2\pi}{3}$ and $\phi=0$ since $\hat{\Phi}_{\frac{2\pi}{3}}$ does not modify $\hat{S}^{z}$. Eventually, in Fig.\ref{fig:sketch} we sketch the emerging spin ordering in the GWC state, as well as its correspondence to the regular AF pattern developing in the SU(2) invariant case.

\subsubsection{Spin-spin correlations in the $\mathbf{q}$-space}
For the sake of completeness we have carried out a complementary analysis in the reciprocal space, which provides indicators of the described effects taking into account the whole volume of the supercell. Namely, we have investigated the spin structure factors defined as
\begin{align}
    \chi^{z}(\mathbf{q})=\frac{1}{N}\sum_{i,j}\text{e}^{i\mathbf{q}\cdot(\mathbf{r}_i-\mathbf{r}_j)}{\mathcal{S}}^{z}_{ij},
\end{align}
and
\begin{align}
    \chi^{x/y}(\mathbf{q})=\frac{1}{N}\sum_{i,j}\text{e}^{i\mathbf{q}\cdot(\mathbf{r}_i-\mathbf{r}_j)}{\mathcal{S}}^{xx/yy}_{ij},
\end{align}

for $\mathbf{q}$ points presented in Fig. \ref{fig:BZ}. 

\begin{figure}
    \centering
    \includegraphics[width=0.8\linewidth]{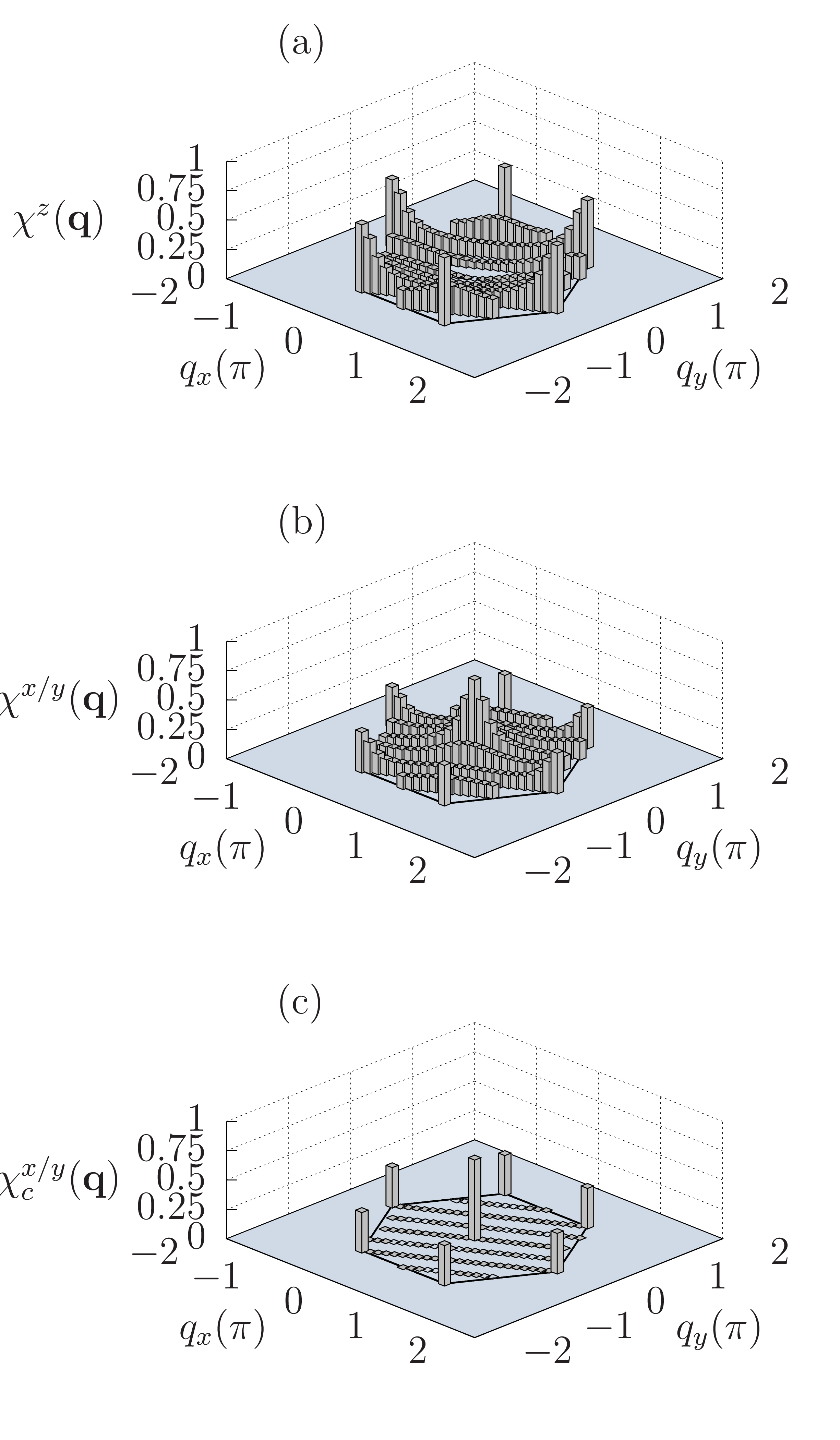}
    \caption{Spin structure factors $\chi^{z}(\mathbf{q})$ and $\chi^{x/y}(\mathbf{q})$  obtained for $V=5|t|$  ((a) and (b) respectively); structure factor $\chi^{x/y}_c(\mathbf{q})$ calculated for  the classical spins alligned in the form as presented in the left panel of Fig.\ref{fig:sketch}, normalized in such way that $\chi^{x/y}_c(\mathbf{\Gamma})$ equals $\chi^{x/y}(\mathbf{\Gamma})$ given in (b).}
    \label{fig:sq}
\end{figure}
In Figs.\ref{fig:sq}(a-b) we explicitly show the spin structure factors resulting from the calculation carried out for $V=5|t|$. The well-pronounced peaks at $\mathbf{K}$ ($\mathbf{K}'$) in $\chi^z$ [Fig.\ref{fig:sq}a] originate from the out-of-plane antiferromagnetic order that develops in the $(110)$ state. However, the landscape of $\chi^{x/y}$ is more complex as is supposed based on the analysis provided in the real space picture. That is, one identifies the peak in $\mathbf{\Gamma}$ that is of greater magnitude than the peaks in the corners of the Brillouin zone ($\mathbf{K}$ and $\mathbf{K}'$ points). However, this behavior in $\chi^{x/y}$  directly corresponds to the order pictured in Fig.\ref{fig:sketch}. To prove it, we computed the spin structure factor $\chi_c^{x/y}$ for the in-plane components, taking the \emph{classical} spins alligned as shown in Fig.\ref{fig:sketch}. As comes from Fig.\ref{fig:sq}c where we plot its distribution in the $\mathbf{q}$-space, the resulting structure of peaks is very similar to that observed in Fig.\ref{fig:sq}b. Indeed, we also find that $\chi_c^{x/y}(\mathbf{\Gamma})/\chi_c^{x/y}(\mathbf{K})=2$, while for the data presented in Fig.\ref{fig:sq}b it is $\approx2.1$, therefore both the positions of the peaks and the relative magnitudes among them agree with the assumed order of the \emph{classical} spins. 

\begin{figure}
    \centering
    \includegraphics[width=0.8\linewidth]{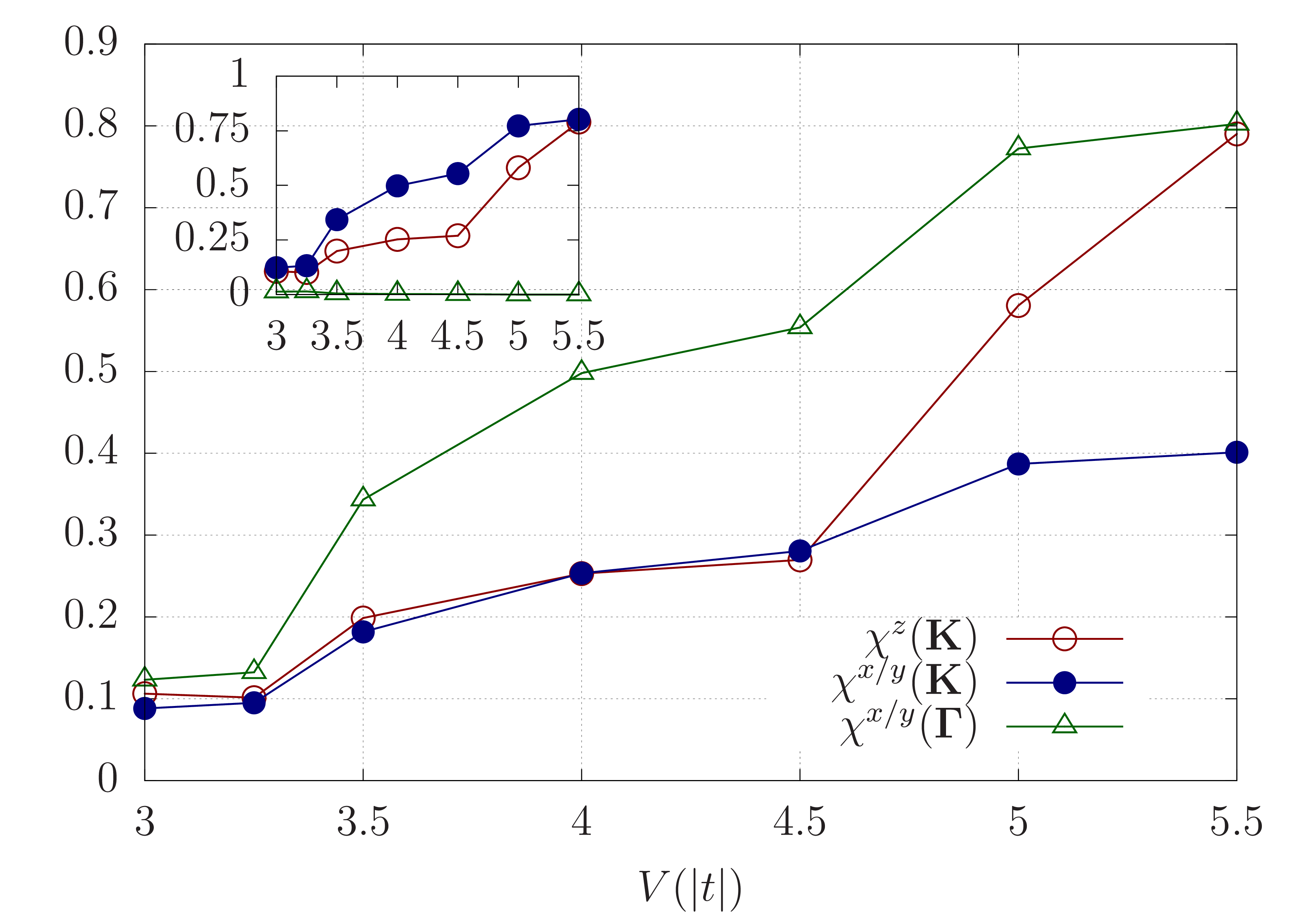}
    \caption{Values of $\chi^z$ and $\chi^{x/y}$ at $\mathbf{q}=\mathbf{\Gamma},\mathbf{K}$ as a function of $V$. In the inset we present the corresponding values when spin-spin correlations where transformed into the case $\phi=0$, one finds that in this case the peak at $\mathbf{\Gamma}$ in $\chi^{x/y}$ vanishes. }
    \label{fig:spinfinal}
\end{figure}

The identified pattern in the $x-y$ plane is robust within the whole range of $V$ in which GWC exists, as comes from Fig.\ref{fig:spinfinal} where we plot the values of $\chi^z$ and $\chi^{x/y}$ at $\mathbf{q}=\mathbf{\Gamma},\mathbf{K}$ as a function of $V$.
The magnitude of the peaks at both $\mathbf{K}$ and $\mathbf{\Gamma}$ starts to increase when $V\gtrsim3.5|t|$, that is, at $V$ for which we estimate the opening of the gap in the $(110)$ phase. In-plane ferromagnetic correlations are dominant up to $V\approx5.5|t|$ for which $\chi^z(\mathbf{K})$ attains a similar magnitude. However, precisely $\chi^z(\mathbf{K})\approx\chi_{x/y}(\mathbf{\Gamma})$ for $V\lesssim4.5|t|$, at $V\approx4.5|t|$ the out-of-plane antiferromagnetic correlations are clearly enhanced.  

Eventually, we conclude that on the basis of the analysis performed for spin-spin correlations $60^{\circ}-$canted  ferromagnetic order develops in the $x-y$ plane while in the out-of-plane direction antiferromagnetic order emerges in the considered model of GWC. Also, it is worth highlighting that the identified order corresponds directly to the purely antiferromgnetic pattern appearing in the SU(2) invariant version of the model (see the inset in Fig.\ref{fig:spinfinal}).

\section{Summary}
In this paper, we have analyzed the minimal model of WSe$_2$/WS$_2$ heterobilayer at fractional filling $2/3$ for which the existence of a generalized Wigner crystal has been experimentally reported. Our approach is based on the extended Hubbard Hamiltonian on a triangular lattice, and the nearest-neighbor Coulomb repulsion $V$ has been treated as a free parameter. Essentially, the non-interacting (single-particle) part of the Hamiltonian has been equipped with complex-valued hopping amplitudes, reproducing the spin-valley splitting which is an inherent feature of this system. The Density Matrix Renormalization group method has been utilized for finding the approximate ground state and its electronic properties. 

By setting the on-site Hubbard interaction $U=15|t|$ and the intersite repulsion in the range $5.5|t|\gtrsim V\gtrsim 3.5|t|$, we were able to reproduce the formation of a generalized Wigner crystal, which is characterized by the $(110)$ sublattice occupation scheme. Namely, the occupied (empty) sites form a honeycomb (triangular) lattice in such a charge pattern. Furthermore, in addition to analyzing the properties of single particles both in real space and momentum, we provide direct evidence of the insulating character of the $(110)$ phase by inspecting charge-charge correlation functions resolved with the Fourier transform at small wave numbers.

Subsequently, we have carried out the analysis of spin properties of the system by studying the spatial behavior of the in- and out-of-plane spin-spin correlation functions separately. Our investigation revealed that in the range of parameters $V$ for which a honeycomb charge pattern appears in the underlying triangular lattice, the strong tendency towards the formation of the out-of-plane antiferromagnetic order concominant with in-plane $60^{\circ}$-canted ferromagnetic order  appears. The former becomes more pronounced when $V$ increases. The identified order is in full agreement with the antiferromagnetic order resulting from the diagonalization of the SU(2) invariant analog of the model considered.

In view of the results obtained here, it can be expected that the magnetic properties of the emerging generalized Wigner crystal in WSe$_2$/WS$_2$ can change with modifications of the experimental setup, such as substrate modifications~\cite{Potasz}, which impact the dielectric function and in turn can change the relative balance between $V$, $U$, and $W$. We hope to see experimental evidence in this regard soon, as well as experimental evidence concerning spin order in these fascinating systems~\cite{Pichler2024}.

\begin{acknowledgments}
This research was funded by National Science Centre, Poland (NCN) according to decision 2021/42/E/ST3/00128. For the purpose of Open Access, the author has applied a CC-BY public copyright licence to any Author Accepted Manuscript (AAM) version arising from this submission.\\

We gratefully acknowledge Poland's high-performance Infrastructure PLGrid ACK Cyfronet AGH for providing computer facilities and support within computational grant no. PLG/2024/017227.

\end{acknowledgments}

\appendix*
\renewcommand{\thefigure}{A\arabic{figure}} 
\setcounter{figure}{0}
\label{appendix}
\section{Error estimation}
\begin{figure}
    \centering
    \includegraphics[width=0.9\linewidth]{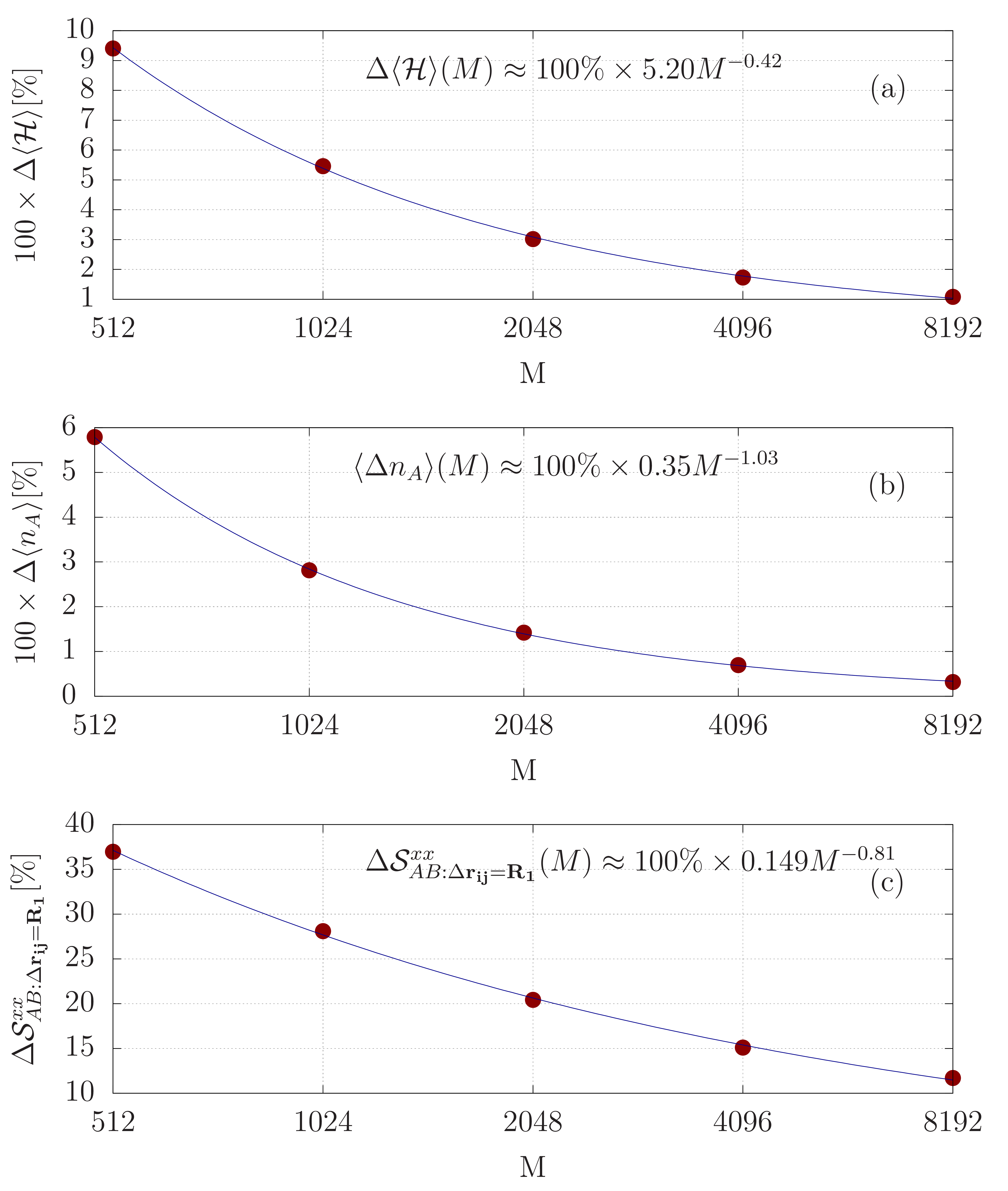}
    \caption{The estimated relative error for $\langle \mathcal{\hat{H}}\rangle(M)$ (a); $\langle n_{A}\rangle(M)$ (b), and $\mathcal{S}^{xx}_{AB,\mathbf{r_{ij}}=\mathbf{R}_1}$ (c). The solid lines correspond to the fits performed. }
    \label{fig:error}
\end{figure}

The system considered here can be regarded as large, specifically in $L_2$, making calculations demanding. In particular, the memory requirements are high; for example, for $M=8192$, which is the maximal link dimension considered here, we need at least $\simeq300GB$ random access memory. The magnitude of $M$ is one of the most important parameters in DMRG calculations in view of the precision of the resulting quantities. Therefore, we performed the convergence analysis on $Q(M)\in\{\langle \mathcal{\hat{H}}\rangle(M),\langle n_{A}\rangle(M), \mathcal{S}^{xx}_{AB,\mathbf{r_{ij}}=\mathbf{R}_1}\}$, that is, on energy, mean value of occupation on $A$ sublattice and the in-plane spin-spin correlation function with the neighboring occupied site belonging to $B$ sublattice. We computed each of the quantities mentioned for bond dimension $M=512,1024,2048,4096,8192$, and the supercell of size $N=6\times24$ at $V=4.5|t|$. Subsequently we fitted obtained data to the algebraic function of form
\begin{align}
    Q(M)=A_{Q}+B_{Q}M^{C_{Q}},
\end{align}
where $A_{Q}\equiv Q(M\rightarrow\infty)$, $B_Q$ and $C_Q$ are parameters to be determined. Eventually, we estimate the relative error given in percents as
\begin{align}
    \Delta Q(M)\approx 100\%\times\Bigg|\frac{A_Q-Q(M)}{A_Q}\Bigg|.
\end{align}
In Fig.\ref{fig:error} we present the resulting estimations. We find that both energy and occupations  for the maximal considered $M=8192$ are very close to their asymptotically predicted values, namely the relative errors are smaller than $0.01\%$. The error associated with spin-spin correlations is notably of greater magnitude, that is $\simeq10\%$, however, such a deviation is not supposed to change the presented conclusions qualitatively.

\section{$S_z^{tot}=0$ assumption} 
\begin{figure}
    \centering
    \includegraphics[width=0.9\linewidth]{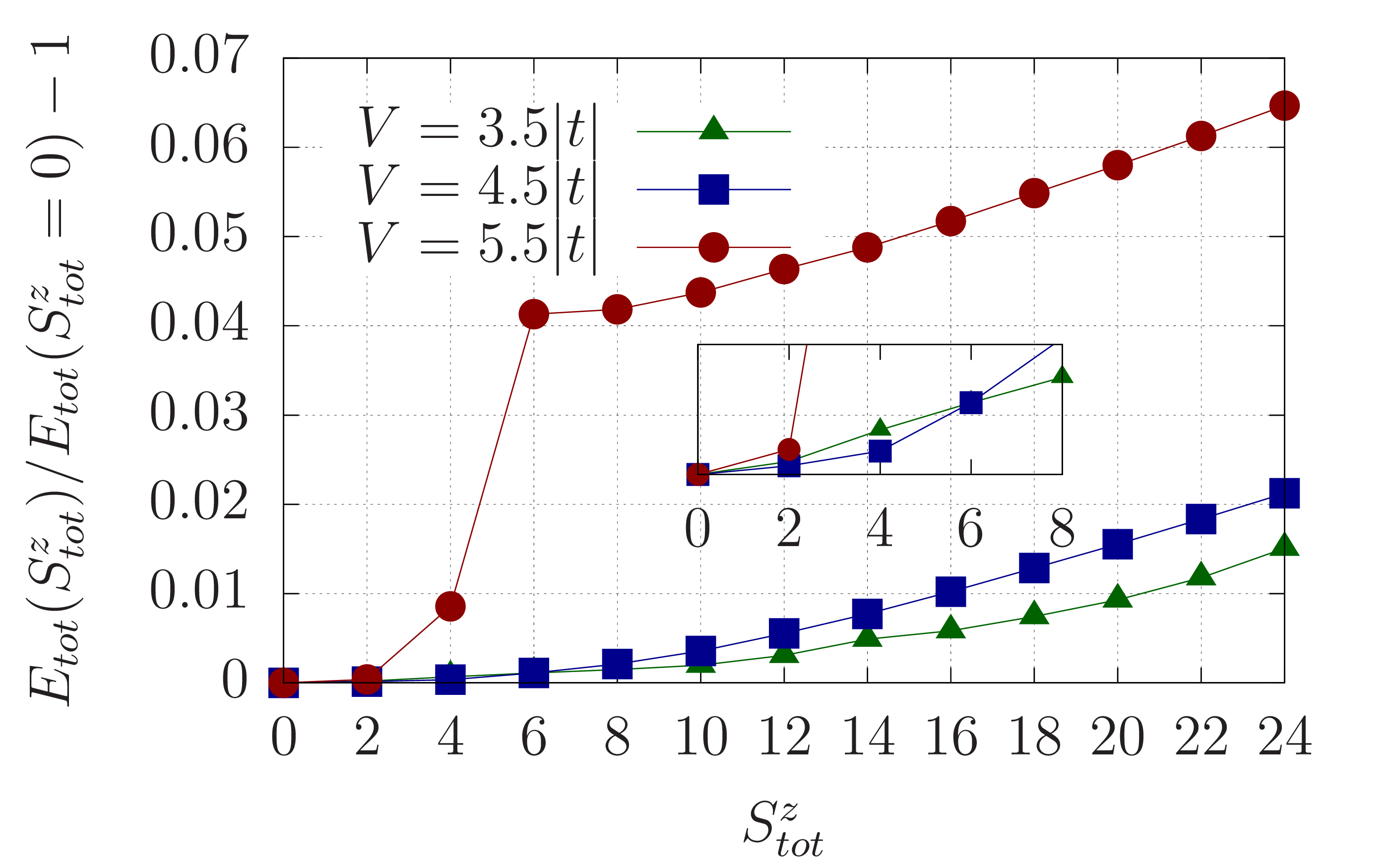}
    \caption{The total energy per lattice site of the  $L_1\times L_2=12\times 6$ supercell at filling $2/3$ for the three values of $V$ representative for $(110)$ state. The lowest energy is obtained for $S_{tot}^z=0$ for each considered $V$. }
    \label{fig:stot}
\end{figure}
To investigate if the $S^z_{tot}=0$ assumption biases our predictions, we performed the set of additional calculations for the supercell of size $L_1\times L_2=12\times6$. We have selected the intersite repulsions $V=\{3.5|t|,4.5|t|,5.5|t|\}$ as a representative for the characterization of $(110)$ phase. We examined the cases for which $S_{tot}^z=\{0,2,4,...24\}$. Note that $S_{tot}^z=24$ refers to the fully polarized system at filling $2/3$ for a cluster of this size. As one can see in Fig.\ref{fig:stot} the minimum of the total energy $E_{tot}$ corresponds to the case where $S_{tot}^{z}=0$ in all three cases. Therefore, we find that the assumption $S_z^{tot}=0$ in the simulations is justified. 

\bibliography{refs.bib}

\begin{thebibliography}{46}%
\makeatletter
\providecommand \@ifxundefined [1]{%
 \@ifx{#1\undefined}
}%
\providecommand \@ifnum [1]{%
 \ifnum #1\expandafter \@firstoftwo
 \else \expandafter \@secondoftwo
 \fi
}%
\providecommand \@ifx [1]{%
 \ifx #1\expandafter \@firstoftwo
 \else \expandafter \@secondoftwo
 \fi
}%
\providecommand \natexlab [1]{#1}%
\providecommand \enquote  [1]{``#1''}%
\providecommand \bibnamefont  [1]{#1}%
\providecommand \bibfnamefont [1]{#1}%
\providecommand \citenamefont [1]{#1}%
\providecommand \href@noop [0]{\@secondoftwo}%
\providecommand \href [0]{\begingroup \@sanitize@url \@href}%
\providecommand \@href[1]{\@@startlink{#1}\@@href}%
\providecommand \@@href[1]{\endgroup#1\@@endlink}%
\providecommand \@sanitize@url [0]{\catcode `\\12\catcode `\$12\catcode `\&12\catcode `\#12\catcode `\^12\catcode `\_12\catcode `\%12\relax}%
\providecommand \@@startlink[1]{}%
\providecommand \@@endlink[0]{}%
\providecommand \url  [0]{\begingroup\@sanitize@url \@url }%
\providecommand \@url [1]{\endgroup\@href {#1}{\urlprefix }}%
\providecommand \urlprefix  [0]{URL }%
\providecommand \Eprint [0]{\href }%
\providecommand \doibase [0]{https://doi.org/}%
\providecommand \selectlanguage [0]{\@gobble}%
\providecommand \bibinfo  [0]{\@secondoftwo}%
\providecommand \bibfield  [0]{\@secondoftwo}%
\providecommand \translation [1]{[#1]}%
\providecommand \BibitemOpen [0]{}%
\providecommand \bibitemStop [0]{}%
\providecommand \bibitemNoStop [0]{.\EOS\space}%
\providecommand \EOS [0]{\spacefactor3000\relax}%
\providecommand \BibitemShut  [1]{\csname bibitem#1\endcsname}%
\let\auto@bib@innerbib\@empty
\bibitem [{\citenamefont {Imada}\ \emph {et~al.}(1998)\citenamefont {Imada}, \citenamefont {Fujimori},\ and\ \citenamefont {Tokura}}]{Imada1998}%
  \BibitemOpen
  \bibfield  {author} {\bibinfo {author} {\bibfnamefont {M.}~\bibnamefont {Imada}}, \bibinfo {author} {\bibfnamefont {A.}~\bibnamefont {Fujimori}},\ and\ \bibinfo {author} {\bibfnamefont {Y.}~\bibnamefont {Tokura}},\ }\bibfield  {title} {\bibinfo {title} {Metal-insulator transitions},\ }\href {https://doi.org/10.1103/RevModPhys.70.1039} {\bibfield  {journal} {\bibinfo  {journal} {Rev. Mod. Phys.}\ }\textbf {\bibinfo {volume} {70}},\ \bibinfo {pages} {1039} (\bibinfo {year} {1998})}\BibitemShut {NoStop}%
\bibitem [{\citenamefont {Cao}\ \emph {et~al.}(2018{\natexlab{a}})\citenamefont {Cao}, \citenamefont {Fatemi}, \citenamefont {Demir}, \citenamefont {Fang}, \citenamefont {Tomarken}, \citenamefont {Luo}, \citenamefont {Sanchez-Yamagishi}, \citenamefont {Watanabe}, \citenamefont {Taniguchi}, \citenamefont {Kaxiras}, \citenamefont {Ashoori},\ and\ \citenamefont {Jarillo-Herrero}}]{Cao2018_1}%
  \BibitemOpen
  \bibfield  {author} {\bibinfo {author} {\bibfnamefont {Y.}~\bibnamefont {Cao}}, \bibinfo {author} {\bibfnamefont {V.}~\bibnamefont {Fatemi}}, \bibinfo {author} {\bibfnamefont {A.}~\bibnamefont {Demir}}, \bibinfo {author} {\bibfnamefont {S.}~\bibnamefont {Fang}}, \bibinfo {author} {\bibfnamefont {S.~L.}\ \bibnamefont {Tomarken}}, \bibinfo {author} {\bibfnamefont {J.~Y.}\ \bibnamefont {Luo}}, \bibinfo {author} {\bibfnamefont {J.~D.}\ \bibnamefont {Sanchez-Yamagishi}}, \bibinfo {author} {\bibfnamefont {K.}~\bibnamefont {Watanabe}}, \bibinfo {author} {\bibfnamefont {T.}~\bibnamefont {Taniguchi}}, \bibinfo {author} {\bibfnamefont {E.}~\bibnamefont {Kaxiras}}, \bibinfo {author} {\bibfnamefont {R.~C.}\ \bibnamefont {Ashoori}},\ and\ \bibinfo {author} {\bibfnamefont {P.}~\bibnamefont {Jarillo-Herrero}},\ }\bibfield  {title} {\bibinfo {title} {Correlated insulator behaviour at half-filling in magic-angle graphene superlattices},\ }\href {https://doi.org/https://doi.org/10.1038/nature26160} {\bibfield  {journal}
  {\bibinfo  {journal} {Nature}\ }\textbf {\bibinfo {volume} {556}},\ \bibinfo {pages} {80} (\bibinfo {year} {2018}{\natexlab{a}})}\BibitemShut {NoStop}%
\bibitem [{\citenamefont {Keimer}\ \emph {et~al.}(2015)\citenamefont {Keimer}, \citenamefont {Kivelson}, \citenamefont {Norman}, \citenamefont {Uchida},\ and\ \citenamefont {Zaanen}}]{Keimer2015}%
  \BibitemOpen
  \bibfield  {author} {\bibinfo {author} {\bibfnamefont {B.}~\bibnamefont {Keimer}}, \bibinfo {author} {\bibfnamefont {S.~A.}\ \bibnamefont {Kivelson}}, \bibinfo {author} {\bibfnamefont {M.~R.}\ \bibnamefont {Norman}}, \bibinfo {author} {\bibfnamefont {S.}~\bibnamefont {Uchida}},\ and\ \bibinfo {author} {\bibfnamefont {J.}~\bibnamefont {Zaanen}},\ }\bibfield  {title} {\bibinfo {title} {From quantum matter to high-temperature superconductivity in copper oxides},\ }\href {https://doi.org/10.1038/nature14165} {\bibfield  {journal} {\bibinfo  {journal} {Nature}\ }\textbf {\bibinfo {volume} {518}},\ \bibinfo {pages} {179} (\bibinfo {year} {2015})}\BibitemShut {NoStop}%
\bibitem [{\citenamefont {Cao}\ \emph {et~al.}(2018{\natexlab{b}})\citenamefont {Cao}, \citenamefont {Fatemi}, \citenamefont {Fang}, \citenamefont {Watanabe}, \citenamefont {Taniguchi}, \citenamefont {Kaxiras},\ and\ \citenamefont {Jarillo-Herrero}}]{Cao2018_2}%
  \BibitemOpen
  \bibfield  {author} {\bibinfo {author} {\bibfnamefont {Y.}~\bibnamefont {Cao}}, \bibinfo {author} {\bibfnamefont {V.}~\bibnamefont {Fatemi}}, \bibinfo {author} {\bibfnamefont {S.}~\bibnamefont {Fang}}, \bibinfo {author} {\bibfnamefont {K.}~\bibnamefont {Watanabe}}, \bibinfo {author} {\bibfnamefont {T.}~\bibnamefont {Taniguchi}}, \bibinfo {author} {\bibfnamefont {E.}~\bibnamefont {Kaxiras}},\ and\ \bibinfo {author} {\bibfnamefont {P.}~\bibnamefont {Jarillo-Herrero}},\ }\bibfield  {title} {\bibinfo {title} {Unconventional superconductivity in magic-angle graphene superlattices},\ }\href {https://doi.org/https://doi.org/10.1038/nature26154} {\bibfield  {journal} {\bibinfo  {journal} {Nature}\ }\textbf {\bibinfo {volume} {556}},\ \bibinfo {pages} {43} (\bibinfo {year} {2018}{\natexlab{b}})}\BibitemShut {NoStop}%
\bibitem [{\citenamefont {Wigner}(1934)}]{Wigner1934}%
  \BibitemOpen
  \bibfield  {author} {\bibinfo {author} {\bibfnamefont {E.}~\bibnamefont {Wigner}},\ }\bibfield  {title} {\bibinfo {title} {On the interaction of electrons in metals},\ }\href {https://doi.org/10.1103/PhysRev.46.1002} {\bibfield  {journal} {\bibinfo  {journal} {Phys. Rev.}\ }\textbf {\bibinfo {volume} {46}},\ \bibinfo {pages} {1002} (\bibinfo {year} {1934})}\BibitemShut {NoStop}%
\bibitem [{\citenamefont {Chiao}(2024)}]{Chiao2024}%
  \BibitemOpen
  \bibfield  {author} {\bibinfo {author} {\bibfnamefont {M.}~\bibnamefont {Chiao}},\ }\bibfield  {title} {\bibinfo {title} {90 years of the wigner crystal},\ }\href {https://doi.org/10.1038/s42254-024-00730-3} {\bibfield  {journal} {\bibinfo  {journal} {Nature Reviews Physics}\ }\textbf {\bibinfo {volume} {6}},\ \bibinfo {pages} {348} (\bibinfo {year} {2024})}\BibitemShut {NoStop}%
\bibitem [{\citenamefont {Tsui}\ \emph {et~al.}(2024)\citenamefont {Tsui}, \citenamefont {He}, \citenamefont {Hu}, \citenamefont {Lake}, \citenamefont {Wang}, \citenamefont {Watanabe}, \citenamefont {Taniguchi}, \citenamefont {Zaletel},\ and\ \citenamefont {Yazdani}}]{Tsui2024}%
  \BibitemOpen
  \bibfield  {author} {\bibinfo {author} {\bibfnamefont {Y.-C.}\ \bibnamefont {Tsui}}, \bibinfo {author} {\bibfnamefont {M.}~\bibnamefont {He}}, \bibinfo {author} {\bibfnamefont {Y.}~\bibnamefont {Hu}}, \bibinfo {author} {\bibfnamefont {E.}~\bibnamefont {Lake}}, \bibinfo {author} {\bibfnamefont {T.}~\bibnamefont {Wang}}, \bibinfo {author} {\bibfnamefont {K.}~\bibnamefont {Watanabe}}, \bibinfo {author} {\bibfnamefont {T.}~\bibnamefont {Taniguchi}}, \bibinfo {author} {\bibfnamefont {M.~P.}\ \bibnamefont {Zaletel}},\ and\ \bibinfo {author} {\bibfnamefont {A.}~\bibnamefont {Yazdani}},\ }\bibfield  {title} {\bibinfo {title} {Direct observation of a magnetic-field-induced wigner crystal},\ }\href {https://doi.org/10.1038/s41586-024-07212-7} {\bibfield  {journal} {\bibinfo  {journal} {Nature}\ }\textbf {\bibinfo {volume} {628}},\ \bibinfo {pages} {287} (\bibinfo {year} {2024})}\BibitemShut {NoStop}%
\bibitem [{\citenamefont {Regan}\ \emph {et~al.}(2020)\citenamefont {Regan}, \citenamefont {Wang}, \citenamefont {Jin}, \citenamefont {Bakti~Utama}, \citenamefont {Gao}, \citenamefont {Wei}, \citenamefont {Zhao}, \citenamefont {Zhao}, \citenamefont {Zhang}, \citenamefont {Yumigeta}, \citenamefont {Blei}, \citenamefont {Carlstr{\"o}m}, \citenamefont {Watanabe}, \citenamefont {Taniguchi}, \citenamefont {Tongay}, \citenamefont {Crommie}, \citenamefont {Zettl},\ and\ \citenamefont {Wang}}]{Regan2020}%
  \BibitemOpen
  \bibfield  {author} {\bibinfo {author} {\bibfnamefont {E.~C.}\ \bibnamefont {Regan}}, \bibinfo {author} {\bibfnamefont {D.}~\bibnamefont {Wang}}, \bibinfo {author} {\bibfnamefont {C.}~\bibnamefont {Jin}}, \bibinfo {author} {\bibfnamefont {M.~I.}\ \bibnamefont {Bakti~Utama}}, \bibinfo {author} {\bibfnamefont {B.}~\bibnamefont {Gao}}, \bibinfo {author} {\bibfnamefont {X.}~\bibnamefont {Wei}}, \bibinfo {author} {\bibfnamefont {S.}~\bibnamefont {Zhao}}, \bibinfo {author} {\bibfnamefont {W.}~\bibnamefont {Zhao}}, \bibinfo {author} {\bibfnamefont {Z.}~\bibnamefont {Zhang}}, \bibinfo {author} {\bibfnamefont {K.}~\bibnamefont {Yumigeta}}, \bibinfo {author} {\bibfnamefont {M.}~\bibnamefont {Blei}}, \bibinfo {author} {\bibfnamefont {J.~D.}\ \bibnamefont {Carlstr{\"o}m}}, \bibinfo {author} {\bibfnamefont {K.}~\bibnamefont {Watanabe}}, \bibinfo {author} {\bibfnamefont {T.}~\bibnamefont {Taniguchi}}, \bibinfo {author} {\bibfnamefont {S.}~\bibnamefont {Tongay}}, \bibinfo {author} {\bibfnamefont {M.}~\bibnamefont
  {Crommie}}, \bibinfo {author} {\bibfnamefont {A.}~\bibnamefont {Zettl}},\ and\ \bibinfo {author} {\bibfnamefont {F.}~\bibnamefont {Wang}},\ }\bibfield  {title} {\bibinfo {title} {Mott and generalized wigner crystal states in wse2/ws2 moir{\'e} superlattices},\ }\href {https://doi.org/10.1038/s41586-020-2092-4} {\bibfield  {journal} {\bibinfo  {journal} {Nature}\ }\textbf {\bibinfo {volume} {579}},\ \bibinfo {pages} {359} (\bibinfo {year} {2020})}\BibitemShut {NoStop}%
\bibitem [{\citenamefont {Xu}\ \emph {et~al.}(2020)\citenamefont {Xu}, \citenamefont {Liu}, \citenamefont {Rhodes}, \citenamefont {Watanabe}, \citenamefont {Taniguchi}, \citenamefont {Hone}, \citenamefont {Elser}, \citenamefont {Mak},\ and\ \citenamefont {Shan}}]{Xu2020}%
  \BibitemOpen
  \bibfield  {author} {\bibinfo {author} {\bibfnamefont {Y.}~\bibnamefont {Xu}}, \bibinfo {author} {\bibfnamefont {S.}~\bibnamefont {Liu}}, \bibinfo {author} {\bibfnamefont {D.~A.}\ \bibnamefont {Rhodes}}, \bibinfo {author} {\bibfnamefont {K.}~\bibnamefont {Watanabe}}, \bibinfo {author} {\bibfnamefont {T.}~\bibnamefont {Taniguchi}}, \bibinfo {author} {\bibfnamefont {J.}~\bibnamefont {Hone}}, \bibinfo {author} {\bibfnamefont {V.}~\bibnamefont {Elser}}, \bibinfo {author} {\bibfnamefont {K.~F.}\ \bibnamefont {Mak}},\ and\ \bibinfo {author} {\bibfnamefont {J.}~\bibnamefont {Shan}},\ }\bibfield  {title} {\bibinfo {title} {Correlated insulating states at fractional fillings of moir{\'e} superlattices},\ }\href {https://doi.org/10.1038/s41586-020-2868-6} {\bibfield  {journal} {\bibinfo  {journal} {Nature}\ }\textbf {\bibinfo {volume} {587}},\ \bibinfo {pages} {214} (\bibinfo {year} {2020})}\BibitemShut {NoStop}%
\bibitem [{\citenamefont {Huang}\ \emph {et~al.}(2021)\citenamefont {Huang}, \citenamefont {Wang}, \citenamefont {Miao}, \citenamefont {Wang}, \citenamefont {Li}, \citenamefont {Lian}, \citenamefont {Taniguchi}, \citenamefont {Watanabe}, \citenamefont {Okamoto}, \citenamefont {Xiao}, \citenamefont {Shi},\ and\ \citenamefont {Cui}}]{Huang2021}%
  \BibitemOpen
  \bibfield  {author} {\bibinfo {author} {\bibfnamefont {X.}~\bibnamefont {Huang}}, \bibinfo {author} {\bibfnamefont {T.}~\bibnamefont {Wang}}, \bibinfo {author} {\bibfnamefont {S.}~\bibnamefont {Miao}}, \bibinfo {author} {\bibfnamefont {C.}~\bibnamefont {Wang}}, \bibinfo {author} {\bibfnamefont {Z.}~\bibnamefont {Li}}, \bibinfo {author} {\bibfnamefont {Z.}~\bibnamefont {Lian}}, \bibinfo {author} {\bibfnamefont {T.}~\bibnamefont {Taniguchi}}, \bibinfo {author} {\bibfnamefont {K.}~\bibnamefont {Watanabe}}, \bibinfo {author} {\bibfnamefont {S.}~\bibnamefont {Okamoto}}, \bibinfo {author} {\bibfnamefont {D.}~\bibnamefont {Xiao}}, \bibinfo {author} {\bibfnamefont {S.-F.}\ \bibnamefont {Shi}},\ and\ \bibinfo {author} {\bibfnamefont {Y.-T.}\ \bibnamefont {Cui}},\ }\bibfield  {title} {\bibinfo {title} {Correlated insulating states at fractional fillings of the ws2/wse2 moir{\'e} lattice},\ }\href {https://doi.org/10.1038/s41567-021-01171-w} {\bibfield  {journal} {\bibinfo  {journal} {Nature Physics}\ }\textbf
  {\bibinfo {volume} {17}},\ \bibinfo {pages} {715} (\bibinfo {year} {2021})}\BibitemShut {NoStop}%
\bibitem [{\citenamefont {Li}\ \emph {et~al.}(2021)\citenamefont {Li}, \citenamefont {Li}, \citenamefont {Regan}, \citenamefont {Wang}, \citenamefont {Zhao}, \citenamefont {Kahn}, \citenamefont {Yumigeta}, \citenamefont {Blei}, \citenamefont {Taniguchi}, \citenamefont {Watanabe}, \citenamefont {Tongay}, \citenamefont {Zettl}, \citenamefont {Crommie},\ and\ \citenamefont {Wang}}]{Li2021}%
  \BibitemOpen
  \bibfield  {author} {\bibinfo {author} {\bibfnamefont {H.}~\bibnamefont {Li}}, \bibinfo {author} {\bibfnamefont {S.}~\bibnamefont {Li}}, \bibinfo {author} {\bibfnamefont {E.~C.}\ \bibnamefont {Regan}}, \bibinfo {author} {\bibfnamefont {D.}~\bibnamefont {Wang}}, \bibinfo {author} {\bibfnamefont {W.}~\bibnamefont {Zhao}}, \bibinfo {author} {\bibfnamefont {S.}~\bibnamefont {Kahn}}, \bibinfo {author} {\bibfnamefont {K.}~\bibnamefont {Yumigeta}}, \bibinfo {author} {\bibfnamefont {M.}~\bibnamefont {Blei}}, \bibinfo {author} {\bibfnamefont {T.}~\bibnamefont {Taniguchi}}, \bibinfo {author} {\bibfnamefont {K.}~\bibnamefont {Watanabe}}, \bibinfo {author} {\bibfnamefont {S.}~\bibnamefont {Tongay}}, \bibinfo {author} {\bibfnamefont {A.}~\bibnamefont {Zettl}}, \bibinfo {author} {\bibfnamefont {M.~F.}\ \bibnamefont {Crommie}},\ and\ \bibinfo {author} {\bibfnamefont {F.}~\bibnamefont {Wang}},\ }\bibfield  {title} {\bibinfo {title} {Imaging two-dimensional generalized wigner crystals},\ }\href
  {https://doi.org/10.1038/s41586-021-03874-9} {\bibfield  {journal} {\bibinfo  {journal} {Nature}\ }\textbf {\bibinfo {volume} {597}},\ \bibinfo {pages} {650} (\bibinfo {year} {2021})}\BibitemShut {NoStop}%
\bibitem [{\citenamefont {Zhou}\ \emph {et~al.}(2024)\citenamefont {Zhou}, \citenamefont {Liang}, \citenamefont {Bi}, \citenamefont {Shi}, \citenamefont {Wang},\ and\ \citenamefont {Li}}]{Zhou2024}%
  \BibitemOpen
  \bibfield  {author} {\bibinfo {author} {\bibfnamefont {H.}~\bibnamefont {Zhou}}, \bibinfo {author} {\bibfnamefont {K.}~\bibnamefont {Liang}}, \bibinfo {author} {\bibfnamefont {L.}~\bibnamefont {Bi}}, \bibinfo {author} {\bibfnamefont {Y.}~\bibnamefont {Shi}}, \bibinfo {author} {\bibfnamefont {Z.}~\bibnamefont {Wang}},\ and\ \bibinfo {author} {\bibfnamefont {S.}~\bibnamefont {Li}},\ }\bibfield  {title} {\bibinfo {title} {Spotlight: Visualization of moir{\'e} quantum phenomena in transition metal dichalcogenide with scanning tunneling microscopy},\ }\href {https://doi.org/10.1021/acsaelm.3c01328} {\bibfield  {journal} {\bibinfo  {journal} {ACS Applied Electronic Materials}\ }\textbf {\bibinfo {volume} {6}},\ \bibinfo {pages} {1530} (\bibinfo {year} {2024})}\BibitemShut {NoStop}%
\bibitem [{\citenamefont {Hubbard}(1978)}]{Hubbard1978}%
  \BibitemOpen
  \bibfield  {author} {\bibinfo {author} {\bibfnamefont {J.}~\bibnamefont {Hubbard}},\ }\bibfield  {title} {\bibinfo {title} {Generalized wigner lattices in one dimension and some applications to tetracyanoquinodimethane (tcnq) salts},\ }\href {https://doi.org/10.1103/PhysRevB.17.494} {\bibfield  {journal} {\bibinfo  {journal} {Phys. Rev. B}\ }\textbf {\bibinfo {volume} {17}},\ \bibinfo {pages} {494} (\bibinfo {year} {1978})}\BibitemShut {NoStop}%
\bibitem [{\citenamefont {Musser}\ and\ \citenamefont {Senthil}(2022)}]{Musser2022}%
  \BibitemOpen
  \bibfield  {author} {\bibinfo {author} {\bibfnamefont {S.}~\bibnamefont {Musser}}\ and\ \bibinfo {author} {\bibfnamefont {T.}~\bibnamefont {Senthil}},\ }\bibfield  {title} {\bibinfo {title} {Metal to wigner-mott insulator transition in two-leg ladders},\ }\href {https://doi.org/10.1103/PhysRevB.106.235148} {\bibfield  {journal} {\bibinfo  {journal} {Phys. Rev. B}\ }\textbf {\bibinfo {volume} {106}},\ \bibinfo {pages} {235148} (\bibinfo {year} {2022})}\BibitemShut {NoStop}%
\bibitem [{\citenamefont {Musser}\ \emph {et~al.}(2022)\citenamefont {Musser}, \citenamefont {Senthil},\ and\ \citenamefont {Chowdhury}}]{Musser2}%
  \BibitemOpen
  \bibfield  {author} {\bibinfo {author} {\bibfnamefont {S.}~\bibnamefont {Musser}}, \bibinfo {author} {\bibfnamefont {T.}~\bibnamefont {Senthil}},\ and\ \bibinfo {author} {\bibfnamefont {D.}~\bibnamefont {Chowdhury}},\ }\bibfield  {title} {\bibinfo {title} {Theory of a continuous bandwidth-tuned wigner-mott transition},\ }\href {https://doi.org/10.1103/PhysRevB.106.155145} {\bibfield  {journal} {\bibinfo  {journal} {Phys. Rev. B}\ }\textbf {\bibinfo {volume} {106}},\ \bibinfo {pages} {155145} (\bibinfo {year} {2022})}\BibitemShut {NoStop}%
\bibitem [{\citenamefont {Wu}\ \emph {et~al.}(2018)\citenamefont {Wu}, \citenamefont {Lovorn}, \citenamefont {Tutuc},\ and\ \citenamefont {MacDonald}}]{Wu2018}%
  \BibitemOpen
  \bibfield  {author} {\bibinfo {author} {\bibfnamefont {F.}~\bibnamefont {Wu}}, \bibinfo {author} {\bibfnamefont {T.}~\bibnamefont {Lovorn}}, \bibinfo {author} {\bibfnamefont {E.}~\bibnamefont {Tutuc}},\ and\ \bibinfo {author} {\bibfnamefont {A.~H.}\ \bibnamefont {MacDonald}},\ }\bibfield  {title} {\bibinfo {title} {Hubbard model physics in transition metal dichalcogenide moir\'e bands},\ }\href {https://doi.org/10.1103/PhysRevLett.121.026402} {\bibfield  {journal} {\bibinfo  {journal} {Phys. Rev. Lett.}\ }\textbf {\bibinfo {volume} {121}},\ \bibinfo {pages} {026402} (\bibinfo {year} {2018})}\BibitemShut {NoStop}%
\bibitem [{\citenamefont {Rademaker}(2022)}]{Rademaker2022}%
  \BibitemOpen
  \bibfield  {author} {\bibinfo {author} {\bibfnamefont {L.}~\bibnamefont {Rademaker}},\ }\bibfield  {title} {\bibinfo {title} {Spin-orbit coupling in transition metal dichalcogenide heterobilayer flat bands},\ }\href {https://doi.org/10.1103/PhysRevB.105.195428} {\bibfield  {journal} {\bibinfo  {journal} {Phys. Rev. B}\ }\textbf {\bibinfo {volume} {105}},\ \bibinfo {pages} {195428} (\bibinfo {year} {2022})}\BibitemShut {NoStop}%
\bibitem [{\citenamefont {Tan}\ \emph {et~al.}(2023)\citenamefont {Tan}, \citenamefont {Tsang}, \citenamefont {Dobrosavljevi\ifmmode~\acute{c}\else \'{c}\fi{}},\ and\ \citenamefont {Rademaker}}]{Tan2023}%
  \BibitemOpen
  \bibfield  {author} {\bibinfo {author} {\bibfnamefont {Y.}~\bibnamefont {Tan}}, \bibinfo {author} {\bibfnamefont {P.~K.~H.}\ \bibnamefont {Tsang}}, \bibinfo {author} {\bibfnamefont {V.}~\bibnamefont {Dobrosavljevi\ifmmode~\acute{c}\else \'{c}\fi{}}},\ and\ \bibinfo {author} {\bibfnamefont {L.}~\bibnamefont {Rademaker}},\ }\bibfield  {title} {\bibinfo {title} {Doping a wigner-mott insulator: Exotic charge orders in transition metal dichalcogenide moir\'e heterobilayers},\ }\href {https://doi.org/10.1103/PhysRevResearch.5.043190} {\bibfield  {journal} {\bibinfo  {journal} {Phys. Rev. Res.}\ }\textbf {\bibinfo {volume} {5}},\ \bibinfo {pages} {043190} (\bibinfo {year} {2023})}\BibitemShut {NoStop}%
\bibitem [{\citenamefont {Motruk}\ \emph {et~al.}(2023)\citenamefont {Motruk}, \citenamefont {Rossi}, \citenamefont {Abanin},\ and\ \citenamefont {Rademaker}}]{Motruk2023}%
  \BibitemOpen
  \bibfield  {author} {\bibinfo {author} {\bibfnamefont {J.}~\bibnamefont {Motruk}}, \bibinfo {author} {\bibfnamefont {D.}~\bibnamefont {Rossi}}, \bibinfo {author} {\bibfnamefont {D.~A.}\ \bibnamefont {Abanin}},\ and\ \bibinfo {author} {\bibfnamefont {L.}~\bibnamefont {Rademaker}},\ }\bibfield  {title} {\bibinfo {title} {Kagome chiral spin liquid in transition metal dichalcogenide moir\'e bilayers},\ }\href {https://doi.org/10.1103/PhysRevResearch.5.L022049} {\bibfield  {journal} {\bibinfo  {journal} {Phys. Rev. Res.}\ }\textbf {\bibinfo {volume} {5}},\ \bibinfo {pages} {L022049} (\bibinfo {year} {2023})}\BibitemShut {NoStop}%
\bibitem [{\citenamefont {Watanabe}\ and\ \citenamefont {Ogata}(2005)}]{Watanabe2005}%
  \BibitemOpen
  \bibfield  {author} {\bibinfo {author} {\bibfnamefont {H.}~\bibnamefont {Watanabe}}\ and\ \bibinfo {author} {\bibfnamefont {M.}~\bibnamefont {Ogata}},\ }\bibfield  {title} {\bibinfo {title} {Charge order and superconductivity in two-dimensional triangular lattice at n=2/3},\ }\href {https://doi.org/10.1143/JPSJ.74.2901} {\bibfield  {journal} {\bibinfo  {journal} {Journal of the Physical Society of Japan}\ }\textbf {\bibinfo {volume} {74}},\ \bibinfo {pages} {2901} (\bibinfo {year} {2005})},\ \Eprint {https://arxiv.org/abs/https://doi.org/10.1143/JPSJ.74.2901} {https://doi.org/10.1143/JPSJ.74.2901} \BibitemShut {NoStop}%
\bibitem [{\citenamefont {Morales-Dur\'an}\ \emph {et~al.}(2023)\citenamefont {Morales-Dur\'an}, \citenamefont {Potasz},\ and\ \citenamefont {MacDonald}}]{Potasz}%
  \BibitemOpen
  \bibfield  {author} {\bibinfo {author} {\bibfnamefont {N.}~\bibnamefont {Morales-Dur\'an}}, \bibinfo {author} {\bibfnamefont {P.}~\bibnamefont {Potasz}},\ and\ \bibinfo {author} {\bibfnamefont {A.~H.}\ \bibnamefont {MacDonald}},\ }\bibfield  {title} {\bibinfo {title} {Magnetism and quantum melting in moir\'e-material wigner crystals},\ }\href {https://doi.org/10.1103/PhysRevB.107.235131} {\bibfield  {journal} {\bibinfo  {journal} {Phys. Rev. B}\ }\textbf {\bibinfo {volume} {107}},\ \bibinfo {pages} {235131} (\bibinfo {year} {2023})}\BibitemShut {NoStop}%
\bibitem [{\citenamefont {Ung}\ \emph {et~al.}(2023)\citenamefont {Ung}, \citenamefont {Lee},\ and\ \citenamefont {Reichman}}]{Ung2023}%
  \BibitemOpen
  \bibfield  {author} {\bibinfo {author} {\bibfnamefont {S.~F.}\ \bibnamefont {Ung}}, \bibinfo {author} {\bibfnamefont {J.}~\bibnamefont {Lee}},\ and\ \bibinfo {author} {\bibfnamefont {D.~R.}\ \bibnamefont {Reichman}},\ }\bibfield  {title} {\bibinfo {title} {Competing generalized wigner crystal states in moir\'e heterostructures},\ }\href {https://doi.org/10.1103/PhysRevB.108.245113} {\bibfield  {journal} {\bibinfo  {journal} {Phys. Rev. B}\ }\textbf {\bibinfo {volume} {108}},\ \bibinfo {pages} {245113} (\bibinfo {year} {2023})}\BibitemShut {NoStop}%
\bibitem [{\citenamefont {Amaricci}\ \emph {et~al.}(2010)\citenamefont {Amaricci}, \citenamefont {Camjayi}, \citenamefont {Haule}, \citenamefont {Kotliar}, \citenamefont {Tanaskovi\ifmmode~\acute{c}\else \'{c}\fi{}},\ and\ \citenamefont {Dobrosavljevi\ifmmode~\acute{c}\else \'{c}\fi{}}}]{Amaricci2010}%
  \BibitemOpen
  \bibfield  {author} {\bibinfo {author} {\bibfnamefont {A.}~\bibnamefont {Amaricci}}, \bibinfo {author} {\bibfnamefont {A.}~\bibnamefont {Camjayi}}, \bibinfo {author} {\bibfnamefont {K.}~\bibnamefont {Haule}}, \bibinfo {author} {\bibfnamefont {G.}~\bibnamefont {Kotliar}}, \bibinfo {author} {\bibfnamefont {D.}~\bibnamefont {Tanaskovi\ifmmode~\acute{c}\else \'{c}\fi{}}},\ and\ \bibinfo {author} {\bibfnamefont {V.}~\bibnamefont {Dobrosavljevi\ifmmode~\acute{c}\else \'{c}\fi{}}},\ }\bibfield  {title} {\bibinfo {title} {Extended hubbard model: Charge ordering and wigner-mott transition},\ }\href {https://doi.org/10.1103/PhysRevB.82.155102} {\bibfield  {journal} {\bibinfo  {journal} {Phys. Rev. B}\ }\textbf {\bibinfo {volume} {82}},\ \bibinfo {pages} {155102} (\bibinfo {year} {2010})}\BibitemShut {NoStop}%
\bibitem [{\citenamefont {White}(1992)}]{White1992}%
  \BibitemOpen
  \bibfield  {author} {\bibinfo {author} {\bibfnamefont {S.~R.}\ \bibnamefont {White}},\ }\bibfield  {title} {\bibinfo {title} {Density matrix formulation for quantum renormalization groups},\ }\href {https://doi.org/10.1103/PhysRevLett.69.2863} {\bibfield  {journal} {\bibinfo  {journal} {Phys. Rev. Lett.}\ }\textbf {\bibinfo {volume} {69}},\ \bibinfo {pages} {2863} (\bibinfo {year} {1992})}\BibitemShut {NoStop}%
\bibitem [{\citenamefont {Schollwöck}(2011)}]{SCHOLLWOCK201196}%
  \BibitemOpen
  \bibfield  {author} {\bibinfo {author} {\bibfnamefont {U.}~\bibnamefont {Schollwöck}},\ }\bibfield  {title} {\bibinfo {title} {The density-matrix renormalization group in the age of matrix product states},\ }\href {https://doi.org/https://doi.org/10.1016/j.aop.2010.09.012} {\bibfield  {journal} {\bibinfo  {journal} {Annals of Physics}\ }\textbf {\bibinfo {volume} {326}},\ \bibinfo {pages} {96} (\bibinfo {year} {2011})},\ \bibinfo {note} {january 2011 Special Issue}\BibitemShut {NoStop}%
\bibitem [{\citenamefont {Catarina}\ and\ \citenamefont {Murta}(2023)}]{Catarina2023}%
  \BibitemOpen
  \bibfield  {author} {\bibinfo {author} {\bibfnamefont {G.}~\bibnamefont {Catarina}}\ and\ \bibinfo {author} {\bibfnamefont {B.}~\bibnamefont {Murta}},\ }\bibfield  {title} {\bibinfo {title} {Density-matrix renormalization group: a pedagogical introduction},\ }\href {https://doi.org/10.1140/epjb/s10051-023-00575-2} {\bibfield  {journal} {\bibinfo  {journal} {The European Physical Journal B}\ }\textbf {\bibinfo {volume} {96}},\ \bibinfo {pages} {111} (\bibinfo {year} {2023})}\BibitemShut {NoStop}%
\bibitem [{\citenamefont {Fishman}\ \emph {et~al.}(2022)\citenamefont {Fishman}, \citenamefont {White},\ and\ \citenamefont {Stoudenmire}}]{itensor1}%
  \BibitemOpen
  \bibfield  {author} {\bibinfo {author} {\bibfnamefont {M.}~\bibnamefont {Fishman}}, \bibinfo {author} {\bibfnamefont {S.~R.}\ \bibnamefont {White}},\ and\ \bibinfo {author} {\bibfnamefont {E.~M.}\ \bibnamefont {Stoudenmire}},\ }\bibfield  {title} {\bibinfo {title} {{The ITensor Software Library for Tensor Network Calculations}},\ }\href {https://doi.org/10.21468/SciPostPhysCodeb.4} {\bibfield  {journal} {\bibinfo  {journal} {SciPost Phys. Codebases}\ ,\ \bibinfo {pages} {4}} (\bibinfo {year} {2022})}\BibitemShut {NoStop}%
\bibitem [{\citenamefont {Shirakawa}\ \emph {et~al.}(2017)\citenamefont {Shirakawa}, \citenamefont {Tohyama}, \citenamefont {Kokalj}, \citenamefont {Sota},\ and\ \citenamefont {Yunoki}}]{Shirakawa2017}%
  \BibitemOpen
  \bibfield  {author} {\bibinfo {author} {\bibfnamefont {T.}~\bibnamefont {Shirakawa}}, \bibinfo {author} {\bibfnamefont {T.}~\bibnamefont {Tohyama}}, \bibinfo {author} {\bibfnamefont {J.}~\bibnamefont {Kokalj}}, \bibinfo {author} {\bibfnamefont {S.}~\bibnamefont {Sota}},\ and\ \bibinfo {author} {\bibfnamefont {S.}~\bibnamefont {Yunoki}},\ }\bibfield  {title} {\bibinfo {title} {Ground-state phase diagram of the triangular lattice hubbard model by the density-matrix renormalization group method},\ }\href {https://doi.org/10.1103/PhysRevB.96.205130} {\bibfield  {journal} {\bibinfo  {journal} {Phys. Rev. B}\ }\textbf {\bibinfo {volume} {96}},\ \bibinfo {pages} {205130} (\bibinfo {year} {2017})}\BibitemShut {NoStop}%
\bibitem [{\citenamefont {Szasz}\ \emph {et~al.}(2020)\citenamefont {Szasz}, \citenamefont {Motruk}, \citenamefont {Zaletel},\ and\ \citenamefont {Moore}}]{Szasz2020}%
  \BibitemOpen
  \bibfield  {author} {\bibinfo {author} {\bibfnamefont {A.}~\bibnamefont {Szasz}}, \bibinfo {author} {\bibfnamefont {J.}~\bibnamefont {Motruk}}, \bibinfo {author} {\bibfnamefont {M.~P.}\ \bibnamefont {Zaletel}},\ and\ \bibinfo {author} {\bibfnamefont {J.~E.}\ \bibnamefont {Moore}},\ }\bibfield  {title} {\bibinfo {title} {Chiral spin liquid phase of the triangular lattice hubbard model: A density matrix renormalization group study},\ }\href {https://doi.org/10.1103/PhysRevX.10.021042} {\bibfield  {journal} {\bibinfo  {journal} {Phys. Rev. X}\ }\textbf {\bibinfo {volume} {10}},\ \bibinfo {pages} {021042} (\bibinfo {year} {2020})}\BibitemShut {NoStop}%
\bibitem [{\citenamefont {Szasz}\ and\ \citenamefont {Motruk}(2021)}]{Szasz2021}%
  \BibitemOpen
  \bibfield  {author} {\bibinfo {author} {\bibfnamefont {A.}~\bibnamefont {Szasz}}\ and\ \bibinfo {author} {\bibfnamefont {J.}~\bibnamefont {Motruk}},\ }\bibfield  {title} {\bibinfo {title} {Phase diagram of the anisotropic triangular lattice hubbard model},\ }\href {https://doi.org/10.1103/PhysRevB.103.235132} {\bibfield  {journal} {\bibinfo  {journal} {Phys. Rev. B}\ }\textbf {\bibinfo {volume} {103}},\ \bibinfo {pages} {235132} (\bibinfo {year} {2021})}\BibitemShut {NoStop}%
\bibitem [{\citenamefont {Tang}\ \emph {et~al.}(2020)\citenamefont {Tang}, \citenamefont {Li}, \citenamefont {Li}, \citenamefont {Xu}, \citenamefont {Liu}, \citenamefont {Barmak}, \citenamefont {Watanabe}, \citenamefont {Taniguchi}, \citenamefont {MacDonald}, \citenamefont {Shan},\ and\ \citenamefont {Mak}}]{Tang2020}%
  \BibitemOpen
  \bibfield  {author} {\bibinfo {author} {\bibfnamefont {Y.}~\bibnamefont {Tang}}, \bibinfo {author} {\bibfnamefont {L.}~\bibnamefont {Li}}, \bibinfo {author} {\bibfnamefont {T.}~\bibnamefont {Li}}, \bibinfo {author} {\bibfnamefont {Y.}~\bibnamefont {Xu}}, \bibinfo {author} {\bibfnamefont {S.}~\bibnamefont {Liu}}, \bibinfo {author} {\bibfnamefont {K.}~\bibnamefont {Barmak}}, \bibinfo {author} {\bibfnamefont {K.}~\bibnamefont {Watanabe}}, \bibinfo {author} {\bibfnamefont {T.}~\bibnamefont {Taniguchi}}, \bibinfo {author} {\bibfnamefont {A.~H.}\ \bibnamefont {MacDonald}}, \bibinfo {author} {\bibfnamefont {J.}~\bibnamefont {Shan}},\ and\ \bibinfo {author} {\bibfnamefont {K.~F.}\ \bibnamefont {Mak}},\ }\bibfield  {title} {\bibinfo {title} {Simulation of hubbard model physics in wse2/ws2 moiré superlattices},\ }\href {https://doi.org/https://doi.org/10.1038/s41586-020-2085-3} {\bibfield  {journal} {\bibinfo  {journal} {Nature}\ }\textbf {\bibinfo {volume} {579}},\ \bibinfo {pages} {353} (\bibinfo {year}
  {2020})}\BibitemShut {NoStop}%
\bibitem [{\citenamefont {Tang}\ \emph {et~al.}(2023)\citenamefont {Tang}, \citenamefont {Su}, \citenamefont {Li}, \citenamefont {Xu}, \citenamefont {Liu}, \citenamefont {Watanabe}, \citenamefont {Taniguchi}, \citenamefont {Hone}, \citenamefont {Jian}, \citenamefont {Xu}, \citenamefont {Mak},\ and\ \citenamefont {Shan}}]{Tang2023}%
  \BibitemOpen
  \bibfield  {author} {\bibinfo {author} {\bibfnamefont {Y.}~\bibnamefont {Tang}}, \bibinfo {author} {\bibfnamefont {K.}~\bibnamefont {Su}}, \bibinfo {author} {\bibfnamefont {L.}~\bibnamefont {Li}}, \bibinfo {author} {\bibfnamefont {Y.}~\bibnamefont {Xu}}, \bibinfo {author} {\bibfnamefont {S.}~\bibnamefont {Liu}}, \bibinfo {author} {\bibfnamefont {K.}~\bibnamefont {Watanabe}}, \bibinfo {author} {\bibfnamefont {T.}~\bibnamefont {Taniguchi}}, \bibinfo {author} {\bibfnamefont {J.}~\bibnamefont {Hone}}, \bibinfo {author} {\bibfnamefont {C.-M.}\ \bibnamefont {Jian}}, \bibinfo {author} {\bibfnamefont {C.}~\bibnamefont {Xu}}, \bibinfo {author} {\bibfnamefont {K.~F.}\ \bibnamefont {Mak}},\ and\ \bibinfo {author} {\bibfnamefont {J.}~\bibnamefont {Shan}},\ }\bibfield  {title} {\bibinfo {title} {Evidence of frustrated magnetic interactions in a wigner--mott insulator},\ }\href {https://doi.org/10.1038/s41565-022-01309-8} {\bibfield  {journal} {\bibinfo  {journal} {Nature Nanotechnology}\ }\textbf {\bibinfo {volume}
  {18}},\ \bibinfo {pages} {233} (\bibinfo {year} {2023})}\BibitemShut {NoStop}%
\bibitem [{\citenamefont {Tocchio}\ \emph {et~al.}(2014)\citenamefont {Tocchio}, \citenamefont {Gros}, \citenamefont {Zhang},\ and\ \citenamefont {Eggert}}]{Tocchio2014}%
  \BibitemOpen
  \bibfield  {author} {\bibinfo {author} {\bibfnamefont {L.~F.}\ \bibnamefont {Tocchio}}, \bibinfo {author} {\bibfnamefont {C.}~\bibnamefont {Gros}}, \bibinfo {author} {\bibfnamefont {X.-F.}\ \bibnamefont {Zhang}},\ and\ \bibinfo {author} {\bibfnamefont {S.}~\bibnamefont {Eggert}},\ }\bibfield  {title} {\bibinfo {title} {Phase diagram of the triangular extended hubbard model},\ }\href {https://doi.org/10.1103/PhysRevLett.113.246405} {\bibfield  {journal} {\bibinfo  {journal} {Phys. Rev. Lett.}\ }\textbf {\bibinfo {volume} {113}},\ \bibinfo {pages} {246405} (\bibinfo {year} {2014})}\BibitemShut {NoStop}%
\bibitem [{\citenamefont {Biborski}(2024)}]{andrzej_biborski_zenodo}%
  \BibitemOpen
  \bibfield  {author} {\bibinfo {author} {\bibfnamefont {A.}~\bibnamefont {Biborski}},\ }\bibfield  {title} {\bibinfo {title} {Extended hubbard model with spin-valley polarization dmrg solution},\ }\bibfield  {journal} {\bibinfo  {journal} {Zenodo}\ }\href {https://doi.org/10.5281/zenodo.13380604} {10.5281/zenodo.13380604} (\bibinfo {year} {2024})\BibitemShut {NoStop}%
\bibitem [{\citenamefont {Spa{\l}ek}\ \emph {et~al.}(2022)\citenamefont {Spa{\l}ek}, \citenamefont {Fidrysiak}, \citenamefont {Zegrodnik},\ and\ \citenamefont {Biborski}}]{Spalek2022}%
  \BibitemOpen
  \bibfield  {author} {\bibinfo {author} {\bibfnamefont {J.}~\bibnamefont {Spa{\l}ek}}, \bibinfo {author} {\bibfnamefont {M.}~\bibnamefont {Fidrysiak}}, \bibinfo {author} {\bibfnamefont {M.}~\bibnamefont {Zegrodnik}},\ and\ \bibinfo {author} {\bibfnamefont {A.}~\bibnamefont {Biborski}},\ }\bibfield  {title} {\bibinfo {title} {Superconductivity in high-tc and related strongly correlated systems from variational perspective: Beyond mean field theory},\ }\href {https://doi.org/https://doi.org/10.1016/j.physrep.2022.02.003} {\bibfield  {journal} {\bibinfo  {journal} {Physics Reports}\ }\textbf {\bibinfo {volume} {959}},\ \bibinfo {pages} {1} (\bibinfo {year} {2022})}\BibitemShut {NoStop}%
\bibitem [{\citenamefont {Brinkman}\ and\ \citenamefont {Rice}(1970)}]{Brinkman1970}%
  \BibitemOpen
  \bibfield  {author} {\bibinfo {author} {\bibfnamefont {W.~F.}\ \bibnamefont {Brinkman}}\ and\ \bibinfo {author} {\bibfnamefont {T.~M.}\ \bibnamefont {Rice}},\ }\bibfield  {title} {\bibinfo {title} {Application of gutzwiller's variational method to the metal-insulator transition},\ }\href {https://doi.org/10.1103/PhysRevB.2.4302} {\bibfield  {journal} {\bibinfo  {journal} {Phys. Rev. B}\ }\textbf {\bibinfo {volume} {2}},\ \bibinfo {pages} {4302} (\bibinfo {year} {1970})}\BibitemShut {NoStop}%
\bibitem [{\citenamefont {Senthil}(2008)}]{Senthil}%
  \BibitemOpen
  \bibfield  {author} {\bibinfo {author} {\bibfnamefont {T.}~\bibnamefont {Senthil}},\ }\bibfield  {title} {\bibinfo {title} {Critical fermi surfaces and non-fermi liquid metals},\ }\href {https://doi.org/10.1103/PhysRevB.78.035103} {\bibfield  {journal} {\bibinfo  {journal} {Phys. Rev. B}\ }\textbf {\bibinfo {volume} {78}},\ \bibinfo {pages} {035103} (\bibinfo {year} {2008})}\BibitemShut {NoStop}%
\bibitem [{\citenamefont {Capello}\ \emph {et~al.}(2005)\citenamefont {Capello}, \citenamefont {Becca}, \citenamefont {Fabrizio}, \citenamefont {Sorella},\ and\ \citenamefont {Tosatti}}]{Becca2005}%
  \BibitemOpen
  \bibfield  {author} {\bibinfo {author} {\bibfnamefont {M.}~\bibnamefont {Capello}}, \bibinfo {author} {\bibfnamefont {F.}~\bibnamefont {Becca}}, \bibinfo {author} {\bibfnamefont {M.}~\bibnamefont {Fabrizio}}, \bibinfo {author} {\bibfnamefont {S.}~\bibnamefont {Sorella}},\ and\ \bibinfo {author} {\bibfnamefont {E.}~\bibnamefont {Tosatti}},\ }\bibfield  {title} {\bibinfo {title} {Variational description of mott insulators},\ }\href {https://doi.org/10.1103/PhysRevLett.94.026406} {\bibfield  {journal} {\bibinfo  {journal} {Phys. Rev. Lett.}\ }\textbf {\bibinfo {volume} {94}},\ \bibinfo {pages} {026406} (\bibinfo {year} {2005})}\BibitemShut {NoStop}%
\bibitem [{\citenamefont {Tocchio}\ \emph {et~al.}(2011)\citenamefont {Tocchio}, \citenamefont {Becca},\ and\ \citenamefont {Gros}}]{Tocchio2011}%
  \BibitemOpen
  \bibfield  {author} {\bibinfo {author} {\bibfnamefont {L.~F.}\ \bibnamefont {Tocchio}}, \bibinfo {author} {\bibfnamefont {F.}~\bibnamefont {Becca}},\ and\ \bibinfo {author} {\bibfnamefont {C.}~\bibnamefont {Gros}},\ }\bibfield  {title} {\bibinfo {title} {Backflow correlations in the hubbard model: An efficient tool for the study of the metal-insulator transition and the large-$u$ limit},\ }\href {https://doi.org/10.1103/PhysRevB.83.195138} {\bibfield  {journal} {\bibinfo  {journal} {Phys. Rev. B}\ }\textbf {\bibinfo {volume} {83}},\ \bibinfo {pages} {195138} (\bibinfo {year} {2011})}\BibitemShut {NoStop}%
\bibitem [{\citenamefont {Tocchio}\ \emph {et~al.}(2020)\citenamefont {Tocchio}, \citenamefont {Montorsi},\ and\ \citenamefont {Becca}}]{Tocchio2020}%
  \BibitemOpen
  \bibfield  {author} {\bibinfo {author} {\bibfnamefont {L.~F.}\ \bibnamefont {Tocchio}}, \bibinfo {author} {\bibfnamefont {A.}~\bibnamefont {Montorsi}},\ and\ \bibinfo {author} {\bibfnamefont {F.}~\bibnamefont {Becca}},\ }\bibfield  {title} {\bibinfo {title} {Magnetic and spin-liquid phases in the frustrated $t\ensuremath{-}{t}^{\ensuremath{'}}$ hubbard model on the triangular lattice},\ }\href {https://doi.org/10.1103/PhysRevB.102.115150} {\bibfield  {journal} {\bibinfo  {journal} {Phys. Rev. B}\ }\textbf {\bibinfo {volume} {102}},\ \bibinfo {pages} {115150} (\bibinfo {year} {2020})}\BibitemShut {NoStop}%
\bibitem [{\citenamefont {Biborski}\ \emph {et~al.}(2024)\citenamefont {Biborski}, \citenamefont {W\'ojcik},\ and\ \citenamefont {Zegrodnik}}]{Biborski2024}%
  \BibitemOpen
  \bibfield  {author} {\bibinfo {author} {\bibfnamefont {A.}~\bibnamefont {Biborski}}, \bibinfo {author} {\bibfnamefont {P.}~\bibnamefont {W\'ojcik}},\ and\ \bibinfo {author} {\bibfnamefont {M.}~\bibnamefont {Zegrodnik}},\ }\bibfield  {title} {\bibinfo {title} {Variational monte carlo approach for the hubbard model applied to twisted bilayer ${\mathrm{wse}}_{2}$ at half-filling},\ }\href {https://doi.org/10.1103/PhysRevB.109.125144} {\bibfield  {journal} {\bibinfo  {journal} {Phys. Rev. B}\ }\textbf {\bibinfo {volume} {109}},\ \bibinfo {pages} {125144} (\bibinfo {year} {2024})}\BibitemShut {NoStop}%
\bibitem [{\citenamefont {Padhi}\ \emph {et~al.}(2018)\citenamefont {Padhi}, \citenamefont {Setty},\ and\ \citenamefont {Phillips}}]{Padhi2018}%
  \BibitemOpen
  \bibfield  {author} {\bibinfo {author} {\bibfnamefont {B.}~\bibnamefont {Padhi}}, \bibinfo {author} {\bibfnamefont {C.}~\bibnamefont {Setty}},\ and\ \bibinfo {author} {\bibfnamefont {P.~W.}\ \bibnamefont {Phillips}},\ }\bibfield  {title} {\bibinfo {title} {Doped twisted bilayer graphene near magic angles: Proximity to wigner crystallization, not mott insulation},\ }\href {https://doi.org/10.1021/acs.nanolett.8b02033} {\bibfield  {journal} {\bibinfo  {journal} {Nano Letters}\ }\textbf {\bibinfo {volume} {18}},\ \bibinfo {pages} {6175} (\bibinfo {year} {2018})}\BibitemShut {NoStop}%
\bibitem [{\citenamefont {Zang}\ \emph {et~al.}(2021)\citenamefont {Zang}, \citenamefont {Wang}, \citenamefont {Cano},\ and\ \citenamefont {Millis}}]{Zang2021}%
  \BibitemOpen
  \bibfield  {author} {\bibinfo {author} {\bibfnamefont {J.}~\bibnamefont {Zang}}, \bibinfo {author} {\bibfnamefont {J.}~\bibnamefont {Wang}}, \bibinfo {author} {\bibfnamefont {J.}~\bibnamefont {Cano}},\ and\ \bibinfo {author} {\bibfnamefont {A.~J.}\ \bibnamefont {Millis}},\ }\bibfield  {title} {\bibinfo {title} {Hartree-fock study of the moir\'e hubbard model for twisted bilayer transition metal dichalcogenides},\ }\href {https://doi.org/10.1103/PhysRevB.104.075150} {\bibfield  {journal} {\bibinfo  {journal} {Phys. Rev. B}\ }\textbf {\bibinfo {volume} {104}},\ \bibinfo {pages} {075150} (\bibinfo {year} {2021})}\BibitemShut {NoStop}%
\bibitem [{\citenamefont {Zang}\ \emph {et~al.}(2022)\citenamefont {Zang}, \citenamefont {Wang}, \citenamefont {Cano}, \citenamefont {Georges},\ and\ \citenamefont {Millis}}]{Zang2022}%
  \BibitemOpen
  \bibfield  {author} {\bibinfo {author} {\bibfnamefont {J.}~\bibnamefont {Zang}}, \bibinfo {author} {\bibfnamefont {J.}~\bibnamefont {Wang}}, \bibinfo {author} {\bibfnamefont {J.}~\bibnamefont {Cano}}, \bibinfo {author} {\bibfnamefont {A.}~\bibnamefont {Georges}},\ and\ \bibinfo {author} {\bibfnamefont {A.~J.}\ \bibnamefont {Millis}},\ }\bibfield  {title} {\bibinfo {title} {Dynamical mean-field theory of moir\'e bilayer transition metal dichalcogenides: Phase diagram, resistivity, and quantum criticality},\ }\href {https://doi.org/10.1103/PhysRevX.12.021064} {\bibfield  {journal} {\bibinfo  {journal} {Phys. Rev. X}\ }\textbf {\bibinfo {volume} {12}},\ \bibinfo {pages} {021064} (\bibinfo {year} {2022})}\BibitemShut {NoStop}%
\bibitem [{\citenamefont {Kiese}\ \emph {et~al.}(2022)\citenamefont {Kiese}, \citenamefont {He}, \citenamefont {Hickey}, \citenamefont {Rubio},\ and\ \citenamefont {Kennes}}]{Kennes2022}%
  \BibitemOpen
  \bibfield  {author} {\bibinfo {author} {\bibfnamefont {D.}~\bibnamefont {Kiese}}, \bibinfo {author} {\bibfnamefont {Y.}~\bibnamefont {He}}, \bibinfo {author} {\bibfnamefont {C.}~\bibnamefont {Hickey}}, \bibinfo {author} {\bibfnamefont {A.}~\bibnamefont {Rubio}},\ and\ \bibinfo {author} {\bibfnamefont {D.~M.}\ \bibnamefont {Kennes}},\ }\bibfield  {title} {\bibinfo {title} {Tmds as a platform for spin liquid physics: A strong coupling study of twisted bilayer wse2},\ }\href {https://doi.org/https://doi.org/10.1063/5.0077901} {\bibfield  {journal} {\bibinfo  {journal} {APL Materials}\ }\textbf {\bibinfo {volume} {10}},\ \bibinfo {pages} {031113} (\bibinfo {year} {2022})}\BibitemShut {NoStop}%
\bibitem [{\citenamefont {Pichler}\ \emph {et~al.}(2024)\citenamefont {Pichler}, \citenamefont {Kadow}, \citenamefont {Kuhlenkamp},\ and\ \citenamefont {Knap}}]{Pichler2024}%
  \BibitemOpen
  \bibfield  {author} {\bibinfo {author} {\bibfnamefont {F.}~\bibnamefont {Pichler}}, \bibinfo {author} {\bibfnamefont {W.}~\bibnamefont {Kadow}}, \bibinfo {author} {\bibfnamefont {C.}~\bibnamefont {Kuhlenkamp}},\ and\ \bibinfo {author} {\bibfnamefont {M.}~\bibnamefont {Knap}},\ }\bibfield  {title} {\bibinfo {title} {Probing magnetism in moir\'e heterostructures with quantum twisting microscopes},\ }\href {https://doi.org/10.1103/PhysRevB.110.045116} {\bibfield  {journal} {\bibinfo  {journal} {Phys. Rev. B}\ }\textbf {\bibinfo {volume} {110}},\ \bibinfo {pages} {045116} (\bibinfo {year} {2024})}\BibitemShut {NoStop}%
\end{thebibliography}%

\end{document}